\documentclass[5p, preprint]{elsarticle}
\usepackage[utf8]{inputenc}
\usepackage{graphicx}
\usepackage{booktabs}
\usepackage{url}
\usepackage{amsmath}
\usepackage[colorlinks=true, allcolors=blue]{hyperref}
\usepackage{lineno}

\journal{Nuclear Instruments and Methods: A}
\begin{document}
\begin{frontmatter}
\author[JLab]{J.~D.~Maxwell}\corref{cor1}
\ead{jmaxwell@jlab.org}
\author[JLab]{J.~Brock}
\author[JLab]{C.~Cuevas}
\author[JLab]{H.~Dong}
\author[JLab]{C.~D.~Keith}
\author[ORNL]{J.~J.~Pierce}
\address[JLab]{Thomas Jefferson National Accelerator Facility, Newport News, VA}
\address[ORNL]{Oak Ridge National Laboratory, Oak Ridge, TN}
\cortext[cor1]{Corresponding author}

\title{A New cw-NMR Q-meter for Dynamically Polarized Targets for Particle Physics}

\begin{abstract}
Polarized solid targets produced via Dynamic Nuclear Polarization rely on Continuous-Wave Nuclear Magnetism Resonance measurements to accurately determine the degree of polarization of  bulk samples polarized to nearly 100\%. Since the late 1970's phase sensitive detection methods have been utilized to observe the magnetization of a sample as a small change in inductance under RF excitation near the Larmor frequency of the nuclear species of interest, using a device known as a Q-meter. Liverpool Q-meters, produced in the UK in the 80's and 90's, have been the workhorse devices for these targets for decades, however their age and scarcity has meant new systems are needed. We describe a new cw-NMR system designed and built at Jefferson Lab, which includes a new Q-meter following the Liverpool style to have comparable electronic performance with several improvements to update and adapt the devices for modern use.
\end{abstract}

\end{frontmatter}

\section{Introduction}
Nuclear Magnetic Resonance was initially demonstrated using continuous waves of radio frequency radiation to drive transitions between nuclear Zeeman states, first by Rabi with molecular beams,\cite{rabi_new_1938}, and later in liquids and solids by Bloch and Purcell \cite{bloch_nuclear_1946, purcell_resonance_1946}, resulting in Nobel prizes for all three. Dynamic Nuclear Polarization (DNP) was invented soon after, in part as a technique to enhance NMR signals for easier detection.  While DNP for NMR has experienced a flurry of development and application in recent years~\cite{griffin_high_2010}, from its advent DNP has also been relied upon for nuclear polarization of bulk solids for use as targets in high energy and nuclear physics scattering experiments~\cite{jeffries_history_1991}. The creation of large samples (up to 1500\,g) of material with hyper-polarized protons or deuterons has been crucial to the exploration of spin-dependent properties of matter at particle accelerators across the world.   When utilizing DNP for polarized targets, NMR becomes a necessary gauge of the average polarization of the target material.

Pulsed-NMR techniques, aided by computers powerful enough to perform fast Fourier analysis, have become enormously important and broadly-applied in chemical and biological sciences\,\cite{reif_solid-state_2021}. However, while pulsed-NMR has the great advantage of simultaneously probing a range of frequencies to reveal the composition and structure of a material, continuous-wave NMR (cw-NMR) has remained the preferred method for polarized targets in scattering experiments as it provides an absorption curve that is linearly proportional to the vector polarization, without the need for Fourier transformation. Continuous-wave NMR allows the direct extraction of the polarization of quadrupolar-broadened nuclei such as $^2$H and $^{14}$N from the line shape~\cite{dulya_line-shape_1997}, where in pulsed-NMR portions of the FID signal can be lost, distorting the  Fourier-transformed signal~\cite{reicherz_pulsed_2004, reicherz_recent_2014}. 
The fixed microwave pumping frequency used to drive DNP precludes the scanning of the magnetic field, necessitating instead scanning of the frequency of the sampling circuit.  

In this article, we describe new cw-NMR instrumentation and software developed for polarized target experiments at the Thomas Jefferson National Accelerator Facility (Jefferson Lab, or JLab).  
This effort aimed to replace the Liverpool Q-meter, the RF sampling circuit and phase sensitive millivoltmeter which form the core of a cw-NMR system. The Liverpool Q-meter has been the {\em de facto} standard for NMR in nuclear scattering targets since the 1980s, but is no longer in production and utilizes obsolete components, making maintenance difficult.  The new Q-meter system described here follows the original Liverpool design closely,
but takes advantage of modern components arranged in a more modular design that is intended to both reduce electromagnetic interference and facilitate easier modification and upgrade.
Our system includes a new FPGA-based control and data-acquisition system for the Q-meter that permits communication over ethernet.

The remainder of this article is organized as follows.  A brief review of cw-NMR and the Liverpool Q-meter is given in Sec.~\ref{sec:History}.  The design of the new JLab Q-meter and its accompanying software are described in Sections~\ref{sec:JLab} and \ref{sec:Software}, respectively.  The performance of the system is compared against the Liverpool Q-meter in Sec.~\ref{sec:Performance}, and possible avenues for future modifications and improvements are discussed in Sec.~\ref{sec:Future}.

\section{Q-meters for cw-NMR}
\label{sec:History}
\subsection{cw-NMR}
Continuous-wave NMR measures the vector polarization $P_z$ of nuclei with spin $I$ in polarized target samples,
\begin{equation}
    \label{eq:VectorPolarization}
    P_z = \frac{\langle I_z \rangle}{I}.
\end{equation}
Here $\langle I_z \rangle$ is the expectation value of the spin projection along the $z$-axis, which is specified by a  magnetic field in that direction, $B_z$.  Closely related to the vector polarization are the sample's magnetization $M_z$ and static susceptibility $\chi_0$:
\begin{equation}
    \label{eq:P_M_chi}
    \chi_o = \mu_0 \frac{M_z}{B_z} = n \mu \mu_0 \frac{P_z}{B_z},
\end{equation}
where the number of nuclei of interest is $n$, $\mu$ is their magnetic dipole moment, and $\mu_0$ is the permeability of free space.  In writing Eq.~\ref{eq:P_M_chi} we have assumed the magnetization of the sample due to the nuclear dipoles is small, so that the magnetic field $B$ and magnetic induction $H$ are approximately equivalent.

Magnetic resonance is realized by applying a second, time-varying field $B_x \cos(\omega t)$ perpendicular to the static one.  The behavior of the spins in this field can be described by an AC susceptibility that is frequency dependent and has both real and imaginary parts
\begin{equation}
    \label{ACsusceptibility}
    \chi(\omega) = \chi'(\omega) + i \chi''(\omega).
\end{equation}
Here $\chi'(\omega)$ and $\chi''(\omega)$ respectively describe the dispersion and absorption of the field due to the spins.  When the frequency is near the spins’ Larmor frequency, $\omega_0 = \mu B_z/\hbar I$, the field can drive transitions between adjacent eigenstates of $I_z$, and energy is either absorbed from the field or given to it via stimulated emission.  Using the Kramers-Kronig relations, one can show that the static susceptibility, and therefore the polarization, can be determined from the integrated absorption lineshape \cite{niinikoski_physics_2020},
\begin{equation}
    \label{eq:Absorb}
    P = \frac{2B_z}{\pi \hbar n \mu \mu_0} \int_0^{+\infty} \frac{\chi''(\omega)}{\omega}\; d\omega.
\end{equation}
For our nuclei of interest, $\chi''(\omega)$ is only non-zero near $\omega_0$, so this integral need only be performed in a small range of $\omega$.

In practice the measurement is made by using a coil
of inductance $L_0$ near or embedded within the sample to generate the time-varying RF field and observing the change in its inductance as the field sweeps through resonance.  The spins and coil are coupled via the sample's AC susceptibility and a filling factor $\eta$:
\begin{equation}
\label{eq:FillingFactor}
L_C(\omega)=L_0[1+4\pi\eta\chi(\omega)].
\end{equation}
Because $\eta$ is very sensitive to the geometry of both the coil and target sample, it is customary to calibrate the circuit response using the absorption signal from a known sample polarization.  This calibration is usually made when the spins are in thermal equilibrium with the lattice at temperature $T$, in which case the polarization is given by the Brillioun function,
\begin{equation}
\label{eq:Brillioun}
    P_{\text{TE}}(I) = \frac{2I+1}{2I} \coth \left( \frac{2I+1}{2I}\; x\right) 
	-\frac{1}{2I} \coth \left(\frac{1}{2I}\; x \right)
\end{equation}
where $x \equiv \mu B/kT$.  The most commonly polarized nuclei in dynamically polarized targets are protons (spin-1/2) and deuterons (spin-1).  In these instances, Eq.~\ref{eq:Brillioun} reduces to
\begin{equation}
P_{\text{TE}}\left( \tfrac{1}{2} \right) = \tanh\left(\mu B/kT\right) 
\end{equation}
and 
\begin{equation}
P_{\text{TE}}(1) = \frac{4\tanh\left( \mu B/kT\right)}{3+\tanh^2\left( \mu B/kT\right)} .
\end{equation}
Assuming a linear response of the NMR circuit, the area under the absorption line $A$ scales
directly with the polarization, and we can write
\begin{equation}
    \label{eq:Areas}
    P_{\text{enh}} = \frac{A_{\text{enh}}}{A_{\text{TE}}} P_{\text{TE}} = C_C  A_{\text{enh}},
\end{equation}
where the ratio $C_C=P_{\text{TE}}/A_{\text{TE}}$ is called the calibration constant.

For the electron-scattering experiments performed at Jefferson Lab, the polarized solid targets typically operate at 1~K and 5~T, where protons and deuterons can be polarized to about 90\% and 50\%, respectively.  By comparison, the thermal equilibrium calibrations are usually performed at slightly warmer temperatures of about 1.4~K to benefit from shorter spin-lattice relaxation times.  Here the TE polarizations are only 0.4\% for protons and 0.07\% for deuterons, meaning the response of the NMR electronics must be highly linear over more than two orders of magnitude.
\label{eq:Kramers}

\subsection{Q-meter Basics}
Measuring a change in inductance near $\omega_0$ relies on reducing the effect of the large reactive impedance of the coil $L_0$ via a resonance with some capacitance, tuned so that the circuit resonant frequency matches the nuclear Larmor frequency: $\omega_0=1/\sqrt{L_0 C}$. This tank $LC$ circuit will behave as an oscillator with some underdamping, a loss of energy per cycle. Underdamping can be described using a quality or \textit{Q-factor}, and a change in circuit impedance due to the emission or absorption of energy from nuclear magnetic resonance is measured with a so-called \textit{Q-meter}. 

The coil's impedance should be purely resistive at $\omega_0$, but reactive components grow as the frequency moves away from resonance, creating a characteristic background to a frequency scan called a \textit{Q-curve}\,\cite{court_high_2004}.  While the tuning capacitor may be placed in parallel or in series with the coil, the series-tuned circuit is more common because it is easier to achieve a linear response than the parallel-tuned alternative\;\cite{petricek_linearized_1968}.  In this article we only discuss the series-tuned Q-meter.

Q-meter electronic systems have been used from the beginnings of cw-NMR measurements. At an early conference on polarized targets and sources in Saclay, France in 1966\,\cite{atkinson_technology_1967},  Q-meter systems in use at Rutherford Lab, CERN, Berkley, Harvard, and Argonne were discussed, systems which all made use of tuned resonator circuits observed through audio amplifiers and rectifiers. The NMR signal response being quite small, noise reduction was an immediate priority, prompting the use of lock-in and differential amplifiers to isolate the signal. The frequency integration of the signal was performed using early analog computers, such as the PACE 231R, also used around that time for NASA's space program. 

A major limitation of these early systems was the use of diode demodulators to measure the resonant response. These rectifiers gave the magnitude of the $\chi(\omega)$ signal, but were not exclusively sensitive to $\chi''(\omega)$.  The contribution of the dispersive term is small only as long as the circuit impedance remains constant over the sweep range, which sets a fundamental limitation to the attainable accuracy of magnitude measurements. 

\begin{figure}
	\centering
		\includegraphics[width=\columnwidth]{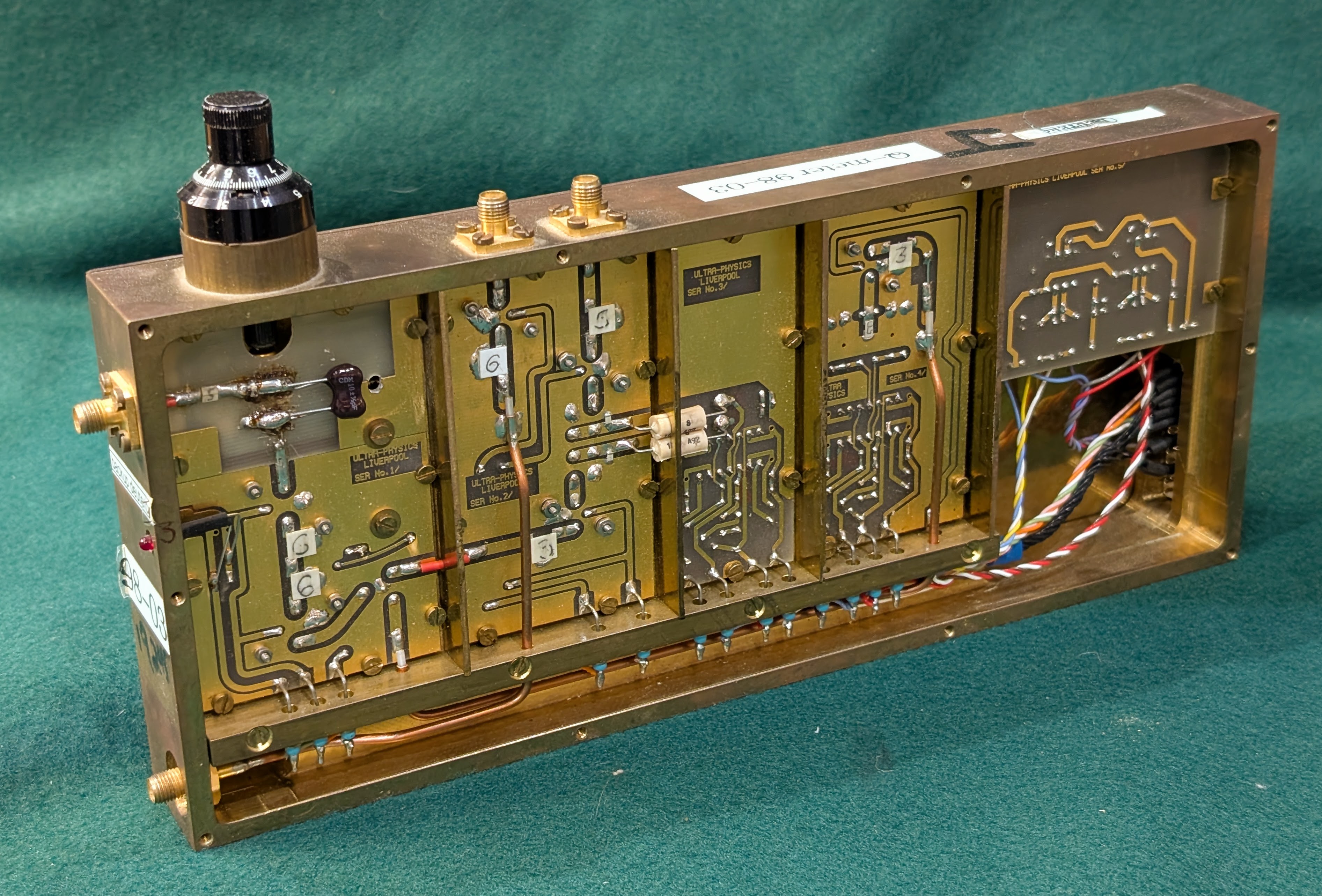}
	\caption{
    A Liverpool Q-meter unit with the cover removed. From left to right are five boards: tank circuit, the mixer, the mixer signal amplification, the diode rectifier, the final amplification. 
    }
	\label{fig:liverpool}
\end{figure}

At the Second Workshop on Polarized Target Materials in 1979, known as the Cosener's House meeting, Court and Gifford introduced the use of phase sensitive detection to isolate the real portion of the Q-meter signal\,\cite{court_proceedings_1980}. Since the complex impedance is related to the inductance as $\sim i L_C(\omega)$, the real portion gives the $\chi''(\omega)$ term. Phase sensitive detection required more complicated electronics using balanced ring modulator (BRM) mixers, but offered improved linearity, reduced noise, and reduced background curvature compared to diode detectors. To implement this scheme for the EMC polarized target, Court and his associates at the University of Liverpool began the small-scale production of a line of cw-NMR systems known as Liverpool Q-meters\,\cite{court_high_1993}, such as the one seen in Figure 
\ref{fig:liverpool}.


\subsection{The Liverpool Q-meter}
\begin{figure}
	\centering
		\includegraphics[width=0.48\columnwidth]{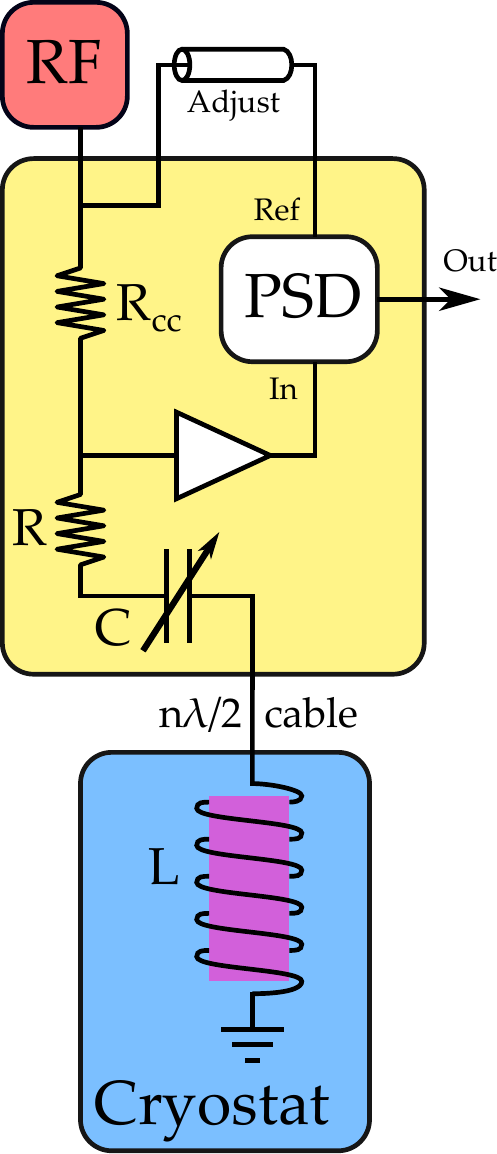}
		\includegraphics[width=0.48\columnwidth]{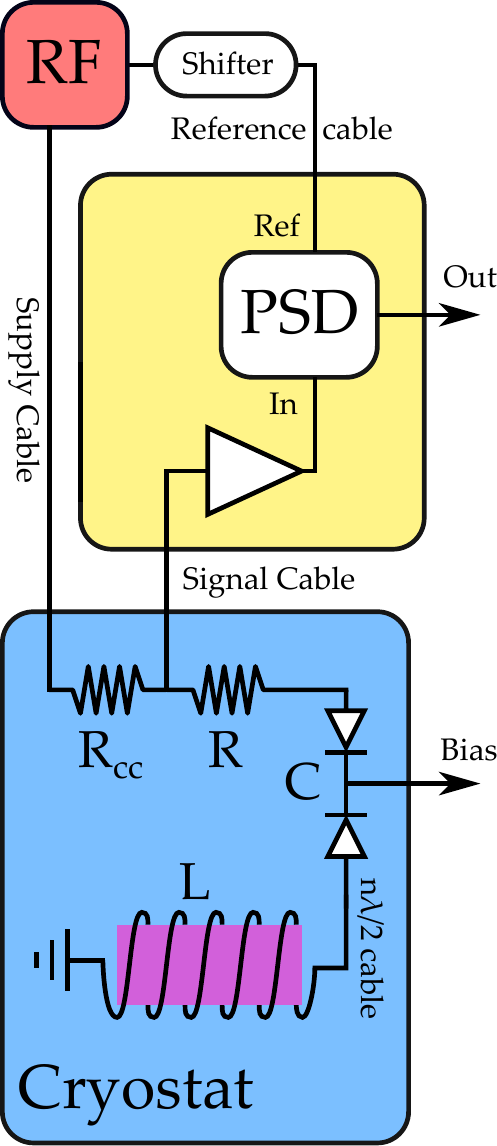}
	\caption{Two simplified diagrams of the Liverpool-style Q-meter circuits built around phase sensitive detectors (PSD).  At left, the tuning capacitance of the LC tank circuit is located outside the target cryostat and is connected to the inductance (NMR coil) using a resonant cable of length $n\lambda/2$. At right, both the capacitance and inductance are inside the cryostat.  In the original Liverpool Q-meter, both the phase and capacitance are adjusted manually.  In the JLab Q-meter, these adjustments are done electronically.}
    
	\label{fig:nmr-circ}
\end{figure}

The Liverpool Q-meter was designed for excellent RF performance with state-of-the-art electronic components for the late 1970's\,\cite{court_development_2004}. During a TE calibration the polarization, and thus signal size, can be over 300 times smaller than that at maximum enhancement, so the Liverpool Q-meter was designed to be linear over a dynamic range of 400:1. The circuit design is at its core relatively simple, as seen in Figure \ref{fig:nmr-circ}, with its heart being the phase sensitive detector, a SRA-1 double-balanced mixer from Mini-circuits. An external RF source is split to drive the LC tank and provide a reference to the mixer. Resistor $R_{CC}$ is designed to be much larger than the maximum impedance of the tank and maintain a constant current condition. The $LC$ circuit consists of a variable capacitor and a coil inductively coupled to the material sample. The voltage across the tank is amplified and split to be sent to the mixer and to a diode rectifier (not shown), providing two final outputs: a phase sensitive signal from the mixer output and a magnitude signal from the diodes.

\subsection{Sample Coil and Resonant Cable}

In DNP applications, embedding the coil in the material requires some part of the circuit enter the refrigerator which maintains near or below 1\;K temperatures. Traditionally, this was accomplished by separating the capacitor, within the Q-meter, and the inductor, within the cryostat as seen at left in Figure \ref{fig:nmr-circ}. In this case the coaxial cable connecting the coil to the Q-meter will likely exceed a resonant length, so the reactive impedance of the cable must be considered. The cable length must be a  half-integer multiple of the RF wavelength so that its impedance at $\omega_0$ is purely resistive. Away from $\omega_0$, this so-called $n\lambda/2$ resonant cable adds impedance, deepening the Q-curve, an effect which worsens at higher values of $n$. This cable should also be chosen to be non-magnetic, minimize heat conduction, and reduce attenuation. PTFE foam insulation is typically chosen, but unfortunately, PTFE has a phase transition near room temperature which can add significant capacitive drift\,\cite{dhawan_precision_1996}.

The coil itself must be designed based on the resonant frequency of the nuclei of interest. In the case of protons at 5\;T, the resonance at 212.9\;MHz means that a very small coil inductance, a single loop about 2\;cm in diameter, is necessary to allow a reasonable tuning capacitance higher than the stray capacitance of the circuit. For the deuteron at 5\;T, the lower Larmor frequency of 32.7\;MHz allows more flexibility in the design of the coil and choice of capacitance.

Some difficulties from added impedance and instability that arise with the use of a long resonant cable can be avoided by moving the capacitance and resistance of the resonant circuit to join the coil within the cryostat~\cite{court_high_2004}. In this scheme, seen at right of Figure \ref{fig:nmr-circ}, two cables connect the tank circuit to the rest of the Q-meter electronics: one supply cable coming from the RF signal generator, and one signal cable returning the tank response. These signal and supply cables must be properly matched, but can be of any length. While this can reduce noise and flatten the Q-curve, it complicates the capacitance tuning and requires the use of radiation-hard components that are temperature stable down at 1\;K. At Jefferson Lab, this scheme has been previously implemented for deuterated materials using a rotary variable capacitor and anticipating the change in capacitance from room temperature down to 1\;K. 

\subsection{Tuning the Q-meter}
Preparing the traditional, resonant-cable Q-meter circuit for experimental use involves three main steps. With the coil already prepared, the capacitor must be adjusted to set the circuit resonance at the nuclear Larmor frequency. The Liverpool Q-meter contains two parallel, variable capacitors to accomplish this, one coarse and one fine. To properly tune the capacitance, the \textit{diode} (magnitude) output of the Q-meter can be plotted versus frequency (on an oscilloscope), and the capacitance shifted until the upward-curving Q-curve is centered at the correct frequency. 
This is ideally done with as short a cable as possible.

Next, the transmission cable between the Q-meter and coil must be confirmed to be an integer multiple of the $\lambda/2$ length. This length depends on the velocity factor the cable.
For the UT-85 semi-rigid cable in frequent use, this length is around 50\;cm for protons and 320\;cm for deuterons at 5\,T.
The cable length is ideally determined electronically, and adjusting with small sections of cable can ensure that the diode Q-curve is still centered at the correct frequency.

Finally, the reference input to the mixer must arrive in-phase with the signal input to ensure that the mixer output is the real portion of the signal. This phase adjustment is made by centering the downward curving Q-curve of the final output of the mixer, or \textit{phase} output, versus frequency.
The Liverpool Q-meter includes two SMA connectors allowing a length of cable to be introduced to delay the reference phase. 


\subsection{Signal Analysis}

Obtaining the polarization from the Q-meter phase signal requires three steps. First the baseline, the background Q-curve that comes from the electronic response of the system, must be removed to leave only the portion of the signal from material polarization. To do this, a Q-curve response is recorded in conditions as close to experimental conditions as possible, but with the polarization removed. This is accomplished by shifting the magnetic field to move the Larmor frequency well out of the frequency sweep range. 

\begin{figure}
	\centering
		\includegraphics[width=\columnwidth]{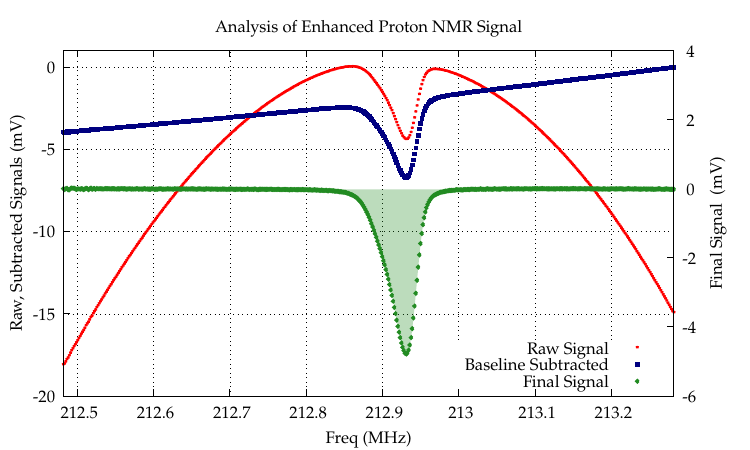}
	\caption{Example analysis steps for a polarized proton signal, showing the raw signal with characteristic ``Q-curve'' background, the signal after subtracting the baseline ``Q-curve'' away (both using left y-axis), and the ``final'' proton polarization signal (using the right y-axis). The area of the shaded green portion is the final output of the signal analysis.}
	\label{fig:te}
\end{figure}

Next, any drift of the Q-curve away from the original baseline must be removed by fitting the portion of the baseline-subtracted signal outside the polarization peak, called the \textit{wings} of the signal, to a polynomial and subtracting that away. The final result should contain only the polarization signal sitting on a flat background for summation, such as seen in Figure \ref{fig:te} for an enhanced proton signal.

\subsection{The Construction of the Liverpool RF Module}
\begin{figure}
	\centering
		\includegraphics[width=\columnwidth]{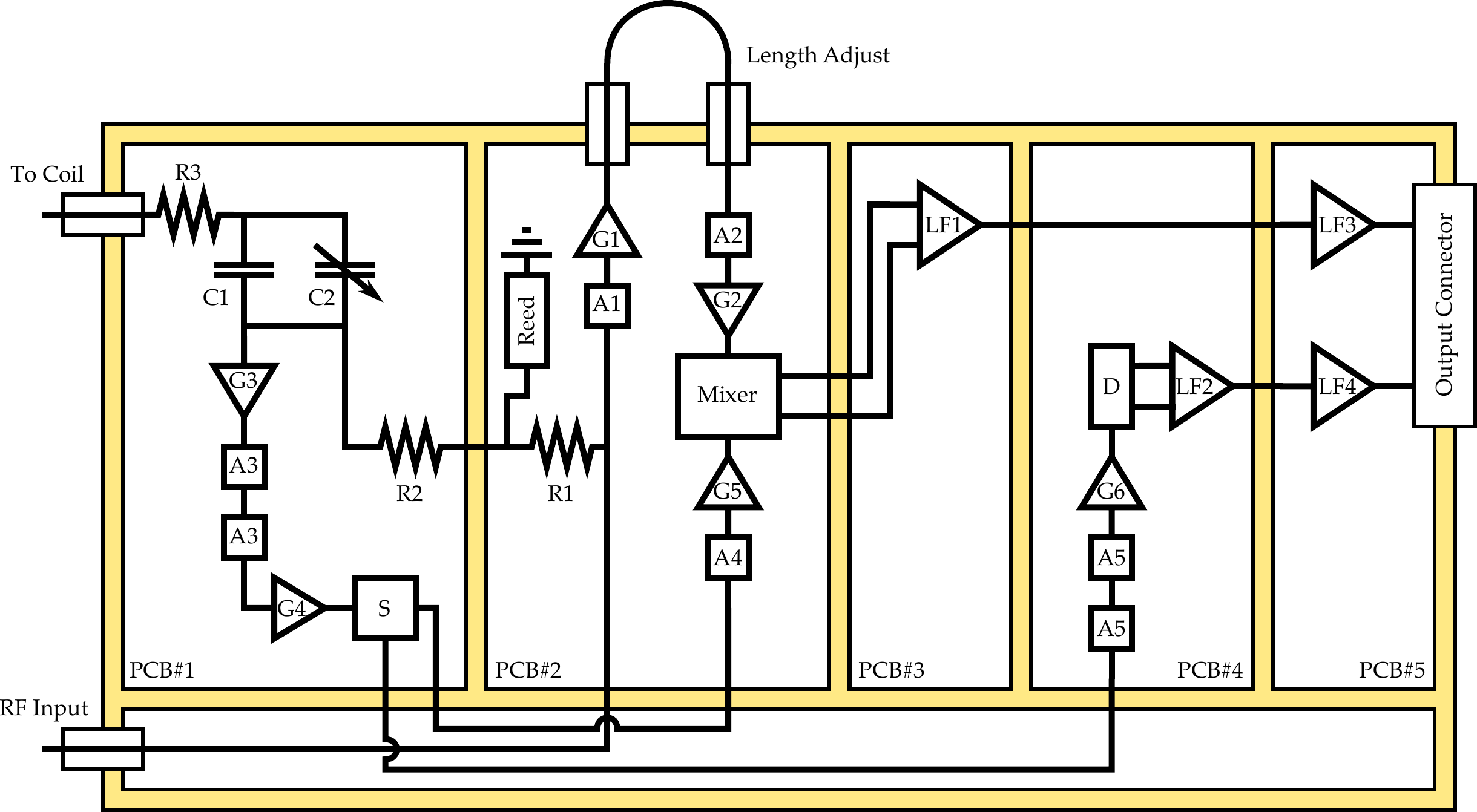}
	\caption{A  schematic layout of the Liverpool RF module configured for use with a resonant cable , orientated left to right as in Figure \ref{fig:liverpool}, showing components and signal routing between the 5 printed circuit boards. Parts denoted ``G'' are attenuators, ``A'' RF amplifiers, ``D'' diodes, ``S'' splitters, and ``LF'' low frequency amplifiers.}
	\label{fig:liv-detail}
\end{figure}

 As seen in Figure \ref{fig:liv-detail}, the Liverpool RF module was housed in a single, machined chassis of gold-plated brass, divided into five, 2-sided, printed circuit boards with shielding partitions between each: three handling high-frequency RF signals, and two low-frequency (Fig.~\ref{fig:liverpool}). The first board contains the tunable capacitance of the tank and stages of RF attenuation and amplification to beat down noise while enhancing the signal size. The second accommodates phase length tuning and holds the RF mixer, which puts out a differential low-frequency signal. This differential signal is amplified on board 3. Board 4 holds the diode bridge, which also puts out a differential signal then amplified on-board. Board 5 has a final stage of low-frequency amplification for the mixer and diode signals.

 The separation of the boards was done quite deliberately to minimize noise in the final phase signal, the most important output of the device. The diode signal is primarily used for tuning and doesn't enter into the error of the polarization. Between each board are brass partitions, and signals pass between the boards through RF chokes. The heavy brass, gold-plated construction of the Q-meter is designed to reduce noise and allow temperature stabilization for the amplifiers.

 The chassis has penetrations for four SMA connectors, and one D-subminiature (Fig.~\ref{fig:liv-detail}). The D-sub carries the differential, low-frequency outputs for the phase and diode signals, as well as power inputs at +24, +15, $\pm$15, and +5 V. The RF signal input is supplied by an external analog signal generator, typically a Rohde \& Schwarz SMT model or similar, and the input is brought into the chassis by a shielded, semi-rigid cable through a separate, shielded channel directly to the mixer board to minimize cross-talk from this largest RF signal in the device.

 While the core electronics of the Liverpool Q-meter are essentially unchanged since the 1980's, the system relied on external control and data acquisition electronics which have varied for different applications. The signal generator provides the necessary RF stimulus and is responsible for frequency modulation to allow integration in frequency. The Liverpool Q-meter was often used with a secondary support board called the Yale card. The Yale card added switchable gain at 1, 20 or 50 times the output of the Q-meter, and included an ADC and DAC to measure and subtract any DC voltage offset in the signal. The output of the Yale card was then sent to a final ADC for measurement, which in the case of the target systems used at Jefferson Lab,  was performed using a National Instruments multi-function board controlled using LabVIEW software.

\section{Designing a New Q-meter}
\label{sec:JLab}
After decades of use in nuclear and particle physics polarized targets, from the landmark EMC experiment to precision spin structure measurements at Jefferson Lab, the Liverpool Q-meter systems began to suffer failures from aging and obsolete components, and multiple groups have seen the need for replacements. At Bochum, a new Q-meter system was designed very much in the Liverpool style\,\cite{herick_neues_2016}. The Bochum Q-meter keeps the heavy brass chassis and partitions, using a single, multi-layer PCB with breaks in the ground plane to divide the high and low frequency sections of the device. Like the Liverpool, the Bochum design makes use of similar RF and DC amplifiers powered by +24, $\pm$15, and +5 V supplies. At Los Alamos National Lab, a rather different Q-meter system was designed from the ground up, feeding the mixer via monolithic microwave integrated circuit amplifiers (MMICs) in a VME form factor allowing for many compact channels\,\cite{mcgaughey_modern_2021}. Despite advancements in electronics over nearly 50 years that brought fast ADC chips capable of digitizing above a giga-sample per second, the frequency down-mixing paradigm of the Liverpool Q-meter still provides excellent noise and sample rate compared to available digital lock-in devices.

Between 1998 and 2012, three separate dynamically polarized target systems were used at Jefferson Lab, on a total of eleven occasions \cite{Keith_polarized_2003, keith_jefferson_2012, maxwell_design_2018, pierce_dynamically_2014}.  Liverpool Q-meters were used for target polarimetry in each instance, but beginning in 2016 the Target and Fast Electronics groups at JLab began an effort to develop new Q-meters for the reasons previously mentioned.  Our primary goal was to match the Liverpool's performance in a modular architecture that would allow easier modification and adaptation to shifting experimental needs.  Additional goals were ergonomic and operational improvements such as voltage-controlled capacitance and phase-shift tuning, and a modernization of the data acquisition and control electronics. The new Q-meter was first utilized with the operation of a new DNP target system for Hall B's CLAS12 detector in a set of experiments in 2022 called Run Group C~\cite{pandey_operation_2024}.

\subsection{Voltage Tuning}
Replacing ``traditional'' tuning using rotary capacitors and varied cable length in favor of voltage tunable options was an important improvement for flexibility and convenience.
Varactor diodes have been used to tune cryogenic tank circuits with Q-meters as far back as 1983\,\cite{veenendaal_frequency-modulated_1983}, and GaAs varactors are both temperature and radiation resistant, making them suitable for use in solid polarized targets. 

Tuning the phase delay to match the arrival of the reference and signal at the mixer is possible with variable delay cables (``trombones'') or with electronic phase shifters. Voltage variable electronic phase shifters, such as those available from Mini-circuits\,\cite{noauthor_phase_nodate}, offer phase tuning of 180$^\circ$ and even 360$^\circ$, albeit in limited frequency ranges.

In practice, we found the added freedom granted by voltage variable tuning of the capacitance and phase can be disorienting to those accustomed to tuning with rotary capacitors and cable lengths, primarily due to the much larger tuning range afforded. Particularly when combined with cold tank circuits using multiple transmission cables in place of a single resonant cable, it is easy find ``false'' tunes. These occur where a Q-curve signal appears to be symmetric in a small frequency window, but what appears is actually a small local minima or maxima due to a mistuned length, possibly inverted by a fully out-of-phase reference. 

\subsubsection{Simulation}
\begin{figure}
	\centering
		\includegraphics[width=\columnwidth]{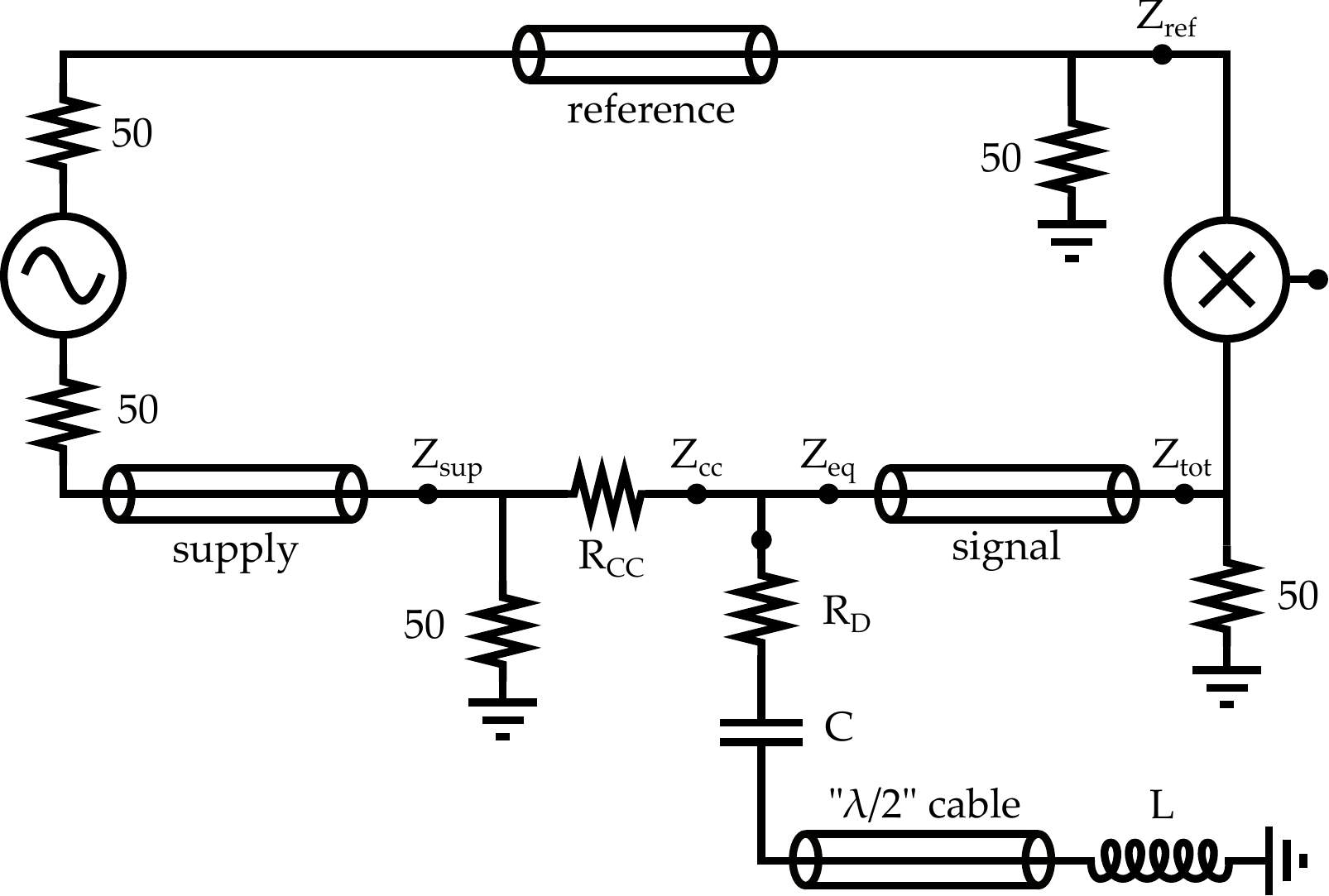}
	\caption{Schematic of a Q-meter circuit as simulated from the input RF at left to the output at the mixer at right. Four variable length transmission lines are shown. The tank circuit consists of $R_d$, $C$, and $L$, and various impedance measurements points in the simulation are shown as $Z$s. Should there be an output from the mixer?}
	\label{fig:sim}
\end{figure}

To better understand the possible tune configurations that come with greatly increased tuning flexibility, we simulated the electronic response of a simplified Q-meter system in Python. This simulation was based initially on Mathcad code by Houlden and Court\,\cite{houlden_nmr_2018}, and was expanded to include transmission line response from a cold tank scheme. Figure \ref{fig:sim} shows the electronic layout of the circuit described by the simulation. The code takes the cable lengths, capacitance and phase tuning as inputs, and calculates the circuit response as both diode and phase outputs.  It is available on GitHub as an interactive notebook\,\cite{maxwell_jdmaxqmetersim_2024}.

\subsection{Q-meter Electronics}
As matching the performance of the Liverpool Q-meter was a chief goal, our electronic design hewed closely to the original. We emulated the steps of amplification and attenuation to keep signal levels close to those in the old boards. As in the Liverpool design, we took care to separate the high and low frequency portions of the circuit into different boards, although we split the circuit slightly differently. We used four boards: a tank board, a reference conditioning board, a high frequency ``mixer'' board, and a low frequency ``amplifier'' board.  For improved shielding and modularity, each board is housed in a dedicated, EMI-shielded enclosure.  These are in turn housed inside a larger aluminum chassis and interconnections between the boards are made with shielded coaxial lines.

\subsubsection{High-Frequency Board}
\begin{figure}
	\centering
		\includegraphics[width= \columnwidth]{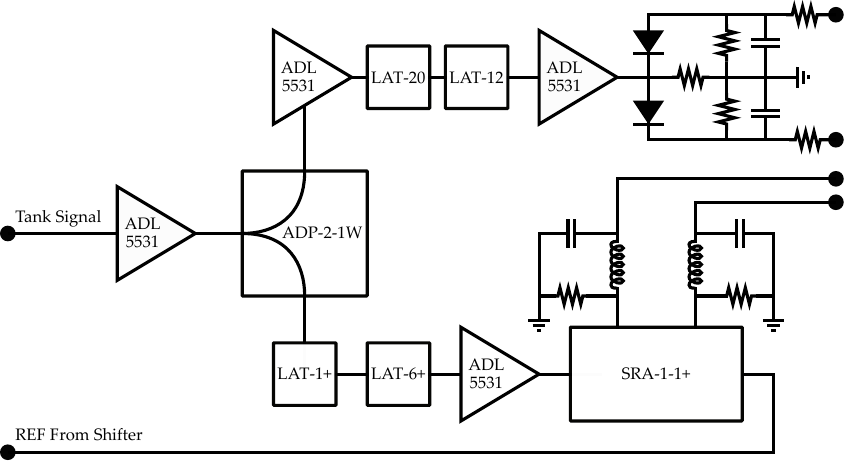}
	\caption{Schematic of the high frequency board electronics, showing amplifiers (ADL), attenuators (LAT), a splitter (ADP) and the mixer itself (SRA).}
	\label{fig:mixer-schem}
\end{figure}

Figure \ref{fig:mixer-schem} shows a schematic of the high-frequency ``mixer'' board. The Liverpool used two types of RF amplifiers, opting for the Watkins-Johnson A71, offering lower gain (18\;dB) with lower noise figure (2.1\;dB) as a first stage of amplification, followed by the A54, offering higher gain (27\;dB) with higher noise figure (4.5\;dB) after an attenuation stage\,\cite{court_liverpool_1980}. In our system we chose one RF amplifier, the Analog Devices ADL5531 20\;dB gain block, which sits between the older amplifiers in terms of gain and still offers an excellent noise figure (2.5\;dB), with improved gain flatness versus temperature and frequency. The linearity of the ADL5531 vastly out-performs the decades-old chips, with an output IP3 of 41 dBm and power at 1 dB compression of 20.6 dB, compared to 19 dBm and 8 dBm for the A54 and 10 dBm and -2.5 dBm for the A71. The ADL5531 operates from 20 to 500 MHz, and so covers both proton and deuteron frequencies at 5\;T. After one amplifier, the signal is split to be sent to the mixer and the diode rectifier chains. The result of each is a differential low-frequency signal which is filtered and sent to the amplifier board via a board-to-board jumper.

The mixer board is a two-layer board with a full ground-plane poured on each side and via-stitching connecting the two ground planes throughout. All components are surface mount, and all signal traces are on the top layer, making them effectively microstrip transmission lines. A few power supply traces jump to the bottom layer where necessary. With few exceptions, a 10\;nF filter capacitor sits between every component. The 5\;V DC power required by the amplifiers enters the board via 3-pin Molex nano-fit connector, and is heavily filtered by 100\;$\mu$H wirewound inductors before being distributed to the amps. The ground plane is unmasked at the edges of the board make thermal and electronic contact with the chassis. Four riser bolts allow the mounting of the amplifier board on top.

\begin{figure}
	\centering
		\includegraphics[width= \columnwidth]{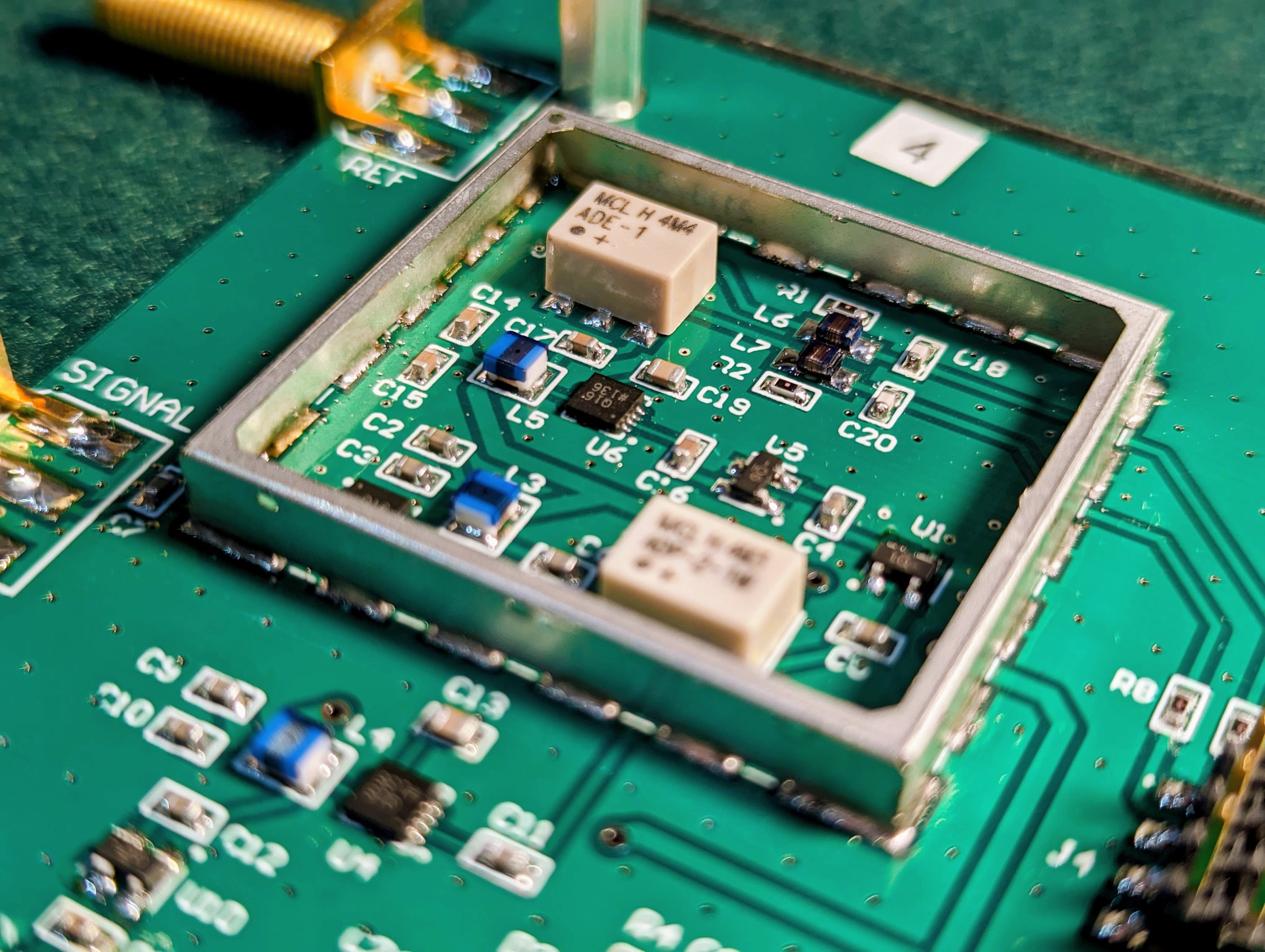}\\
		\includegraphics[width= \columnwidth]{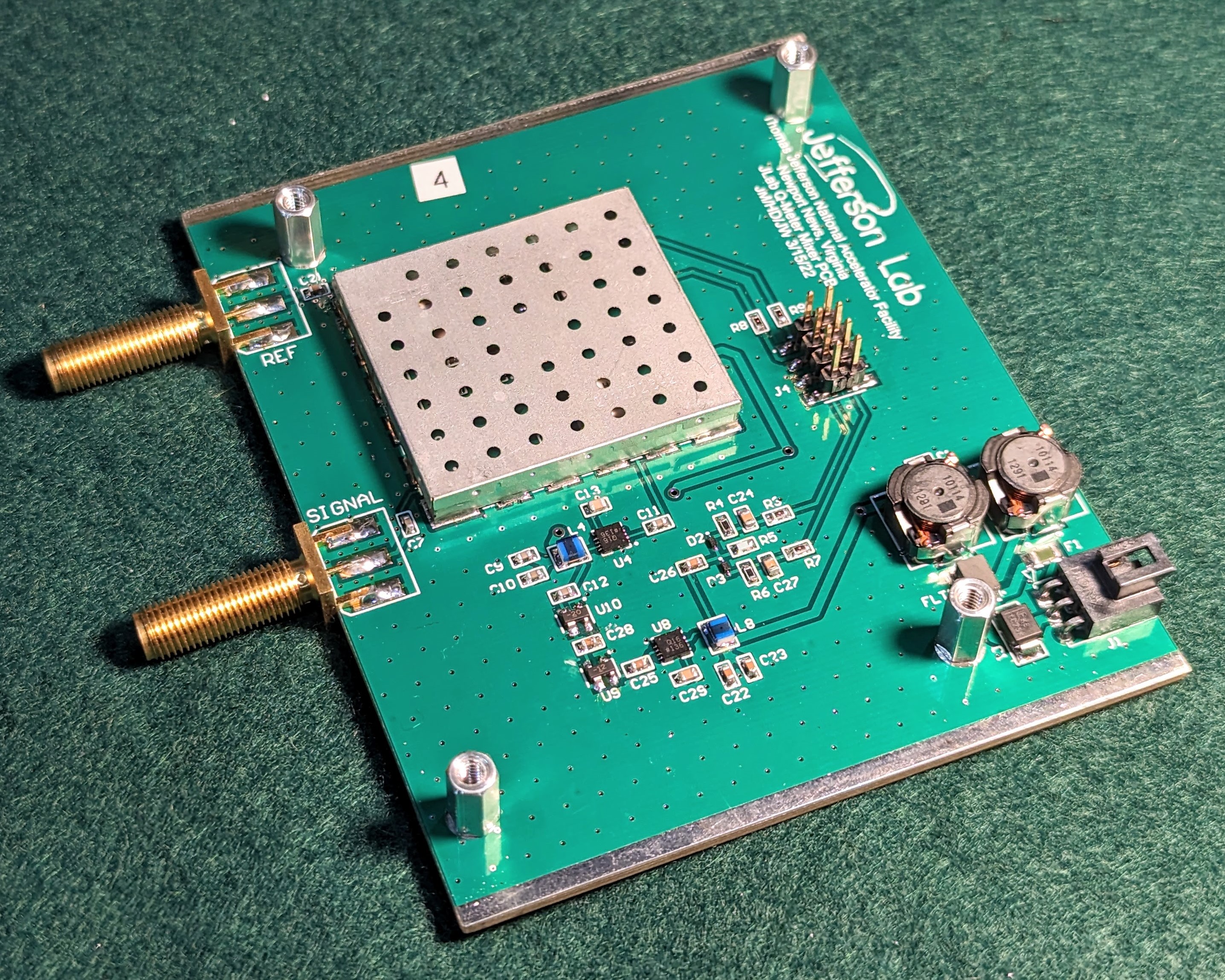}
	\caption{Photographic detail of the mixer electronics within the board level shield with cover removed at top. Full view of the mixer board below.}
	\label{fig:mixer}
\end{figure}

The heart of the Q-meter, the frequency mixer, resides in the innermost layer of the EMI shielding ``onion'': an aluminium board level shield with a removable cover from TE connectivity, seen with cover removed at the top of Figure \ref{fig:mixer}. The reference signal enters the shielded area to immediately enter the mixer, avoiding cross-talk noise. The tank signal is likewise brought directly into the shield, amplified, and split, with the diode side of the split exiting the board shield to be rectified. The mixer itself is an ADE-1+ frequency mixer from Mini-circuits, a surface mount variation of the SRA-1 chosen for the Liverpool back in the 1970's.

\subsubsection{Low-Frequency Board}
\begin{figure}
	\centering
		\includegraphics[width= \columnwidth]{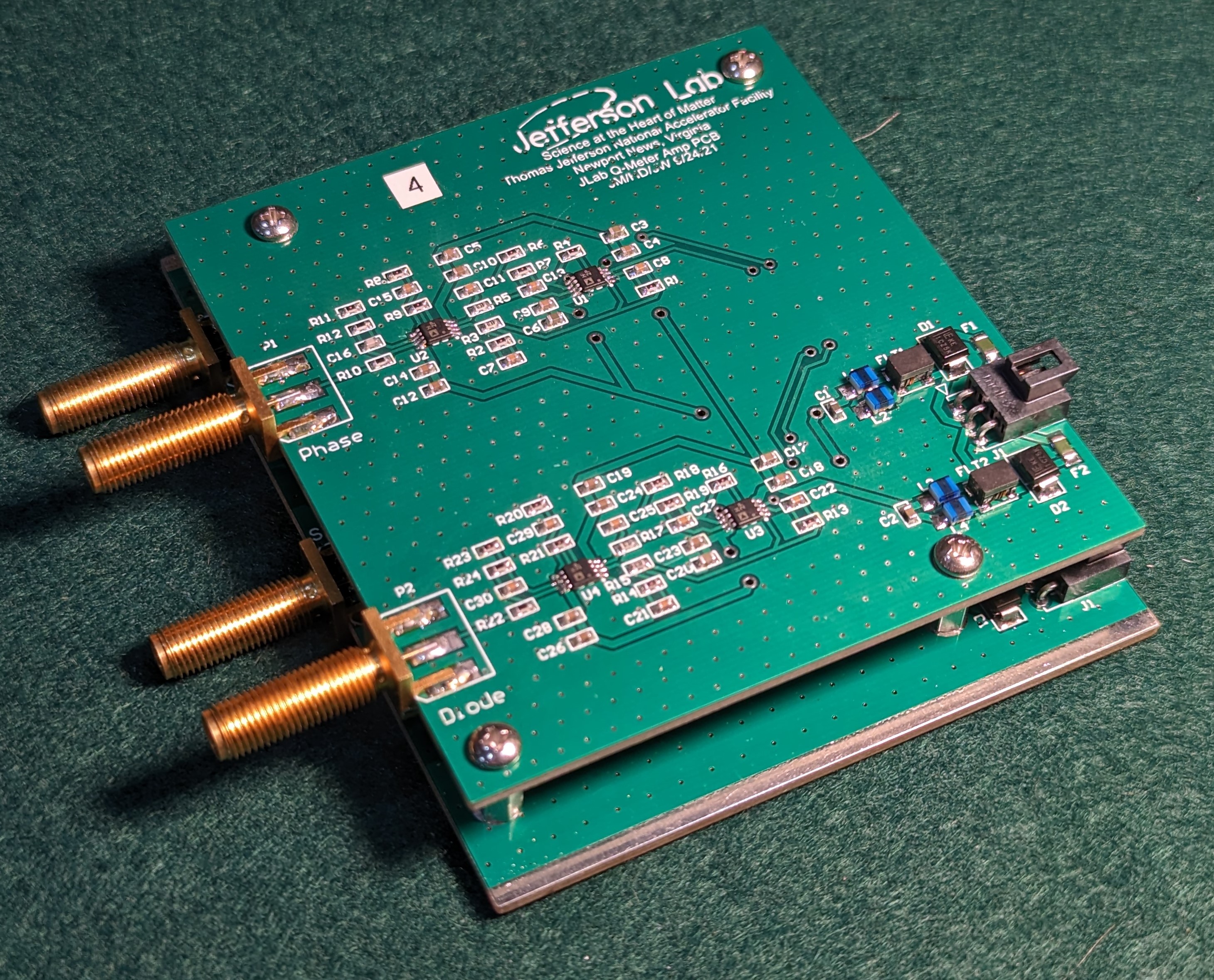}
	\caption{Full view of the amplifier board mounted to the top of the mixer board.}
	\label{fig:amp}
\end{figure}

The down-mixed and rectified signals are brought up through a mezzanine-type board-to-board connection to the low-frequency amplifier board, seen in Figure \ref{fig:amp}. Built with the same techniques as the mixer board, a full ground plane on the bottom layer separates the low-frequency components from the mixer board components below. The board has two identical circuit paths, one each to amplify the mixer output and diode output. Four two-channel Analog Devices ADA4896-2 rail-to-rail amplifiers allow four channels for combining and amplifying each differential signal for output to two SMA connectors.

\subsubsection{Tank Resonator Boards}
The tank circuit was separated out to a small board, allowing it to be located within the cryostat or in a small enclosure near the Q-meter itself. To avoid stray capacitance, these boards were printed without ground planes and with very short traces. 

\begin{figure}
	\centering
		\includegraphics[width= \columnwidth]{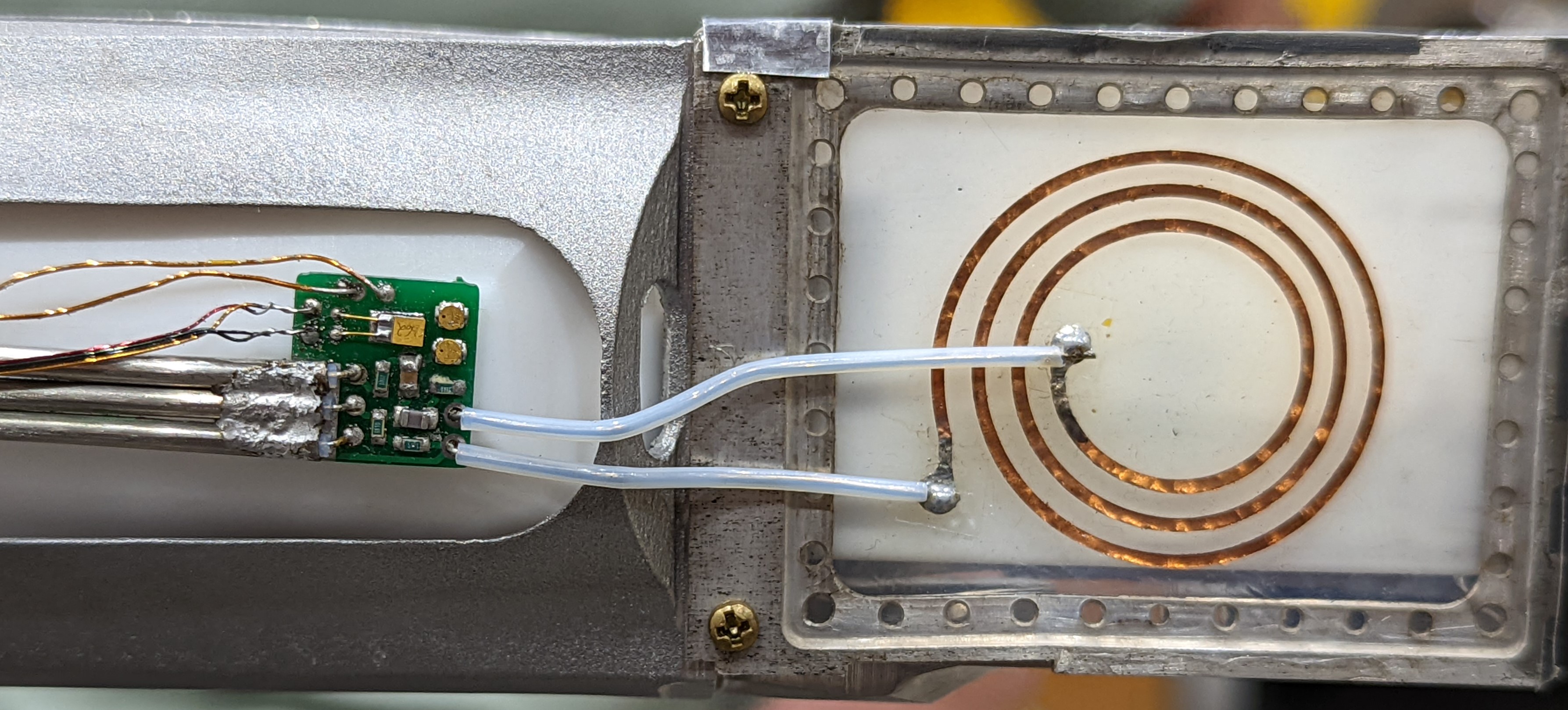}
	\caption{Photograph of the cold tank circuit and deuteron coil mounted in the CLAS12 polarized target refrigerator.  The board is about 1.3\;cm on a side. The coil is laminated in Kel-F film, and the target material cell sits inside the white teflon helium bath behind it, within a couple millimeters of the coil.}
	\label{fig:coldboard}
\end{figure}

In the CLAS12 horizontal DNP refrigerator, the ``cold'' tank circuit board was located just outside the target material's liquid helium bath, as seen in Figure \ref{fig:coldboard}.  Macom MA46 hyperabrupt varactors were used for both cold and warm tank circuits. In our preliminary tests, we found these GaAs varactors to be relatively insensitive to temperature down to about 3\,K.  However, the DNP refrigerator operates at 1~K, and the entire vicinity around the target bath is covered in a superfluid film that created a significant shift in the capacitance of the tank circuit,
perhaps due to the dielectric constant of liquid helium.
We eliminated the film by heating the board to approximately 15\,K using a small metal film resistor on the back of the board and a Cernox thermistor to measure its temperature.
Three coaxial semi-rigid cables were necessary to carry the supply, signal, and varactor bias to the board, as well as 6 additional wires for the heater and thermistor.


The total capacitance ratio of the GaAs tuning varactors we chose was 10 (roughly 2 to 20 pF)\,\cite{noauthor_ma46h204-1056_nodate}, which would be divided in half when used in-series as in Figure \ref{fig:nmr-circ}. However, we created larger tuning range by using two varactors parallel with each other, in-series with a larger static capacitor, doubling the effective capacitance.

\begin{figure}
	\centering
		\includegraphics[width= \columnwidth]{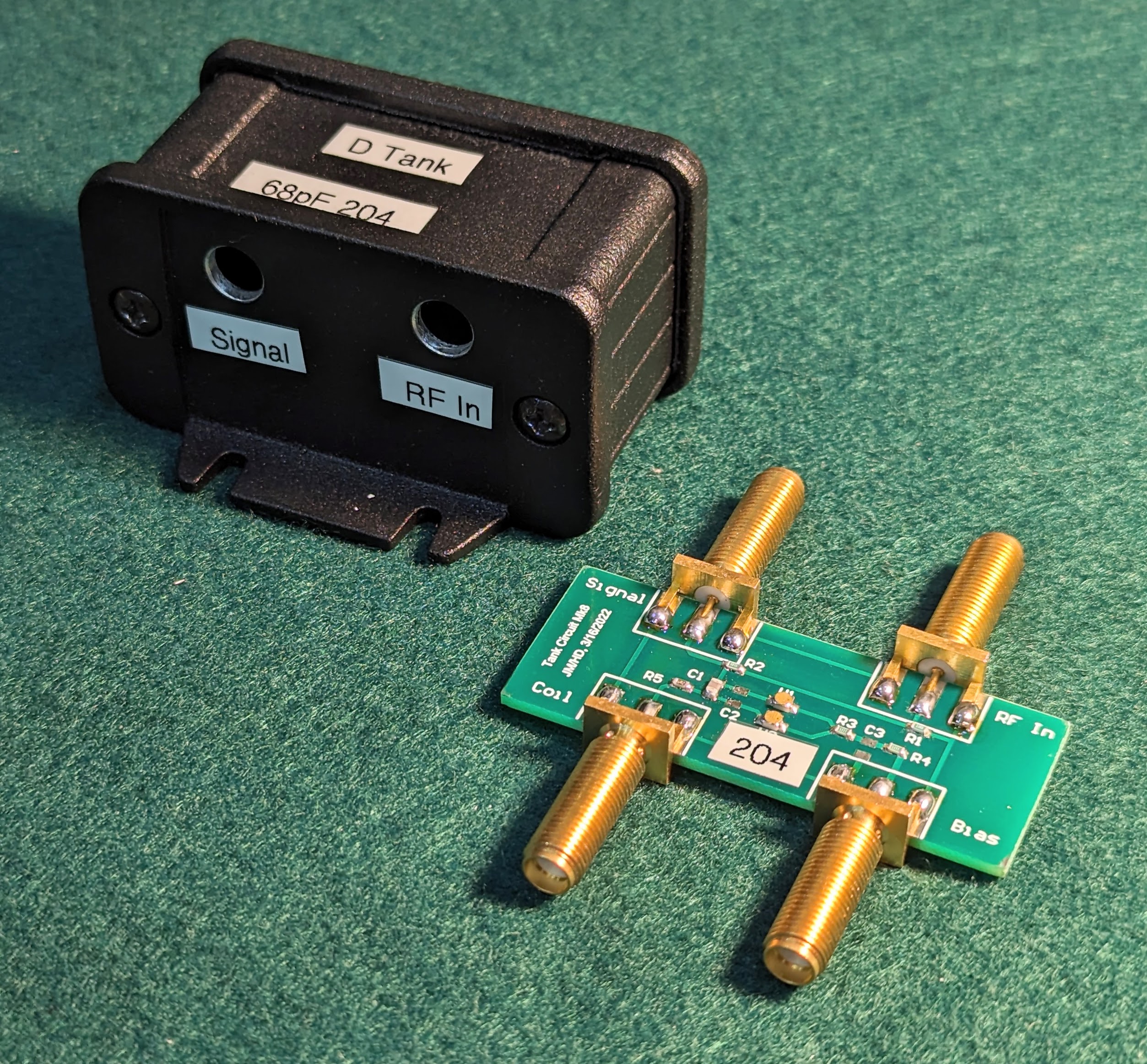}
	\caption{Photograph of the ``warm'' tank board and its enclosure.}
	\label{fig:warmboard}
\end{figure}

In addition to the cold tank board, a more traditional tank circuit board was available for use inside the Q-meter chassis, as seen in Figure \ref{fig:warmboard}. The board used the same varactors, and was connectorized with SMAs for ease of use. The board was housed in a small, shielded enclosure.

\subsubsection{Reference Conditioning Board}
The last board was designed to condition the reference input to the mixer by accommodating an electronic phase shifter and amplifier. Produced in a similar manner to the mixer and amplifier boards, the conditioner board had 2 inputs and 2 outputs. The reference signal from the RF signal generator was brought in and split, with one side exiting the board to supply the tank circuit, and the other being sent through a phase shifter, attenuator and amplifier, then exiting as reference to the mixer. Phase shifters from Mini-circuits are useful in a narrow frequency range, so different phase shift models were necessary. The SCPHS-51+ provided at 360$^\circ$ shift for 32.7\;MHz, and the SPHSA-251+ a 180$^\circ$ shift for 213\;MHz. The PCB was designed to accommodate either component.

\begin{figure}
	\centering
		\includegraphics[width= \columnwidth]{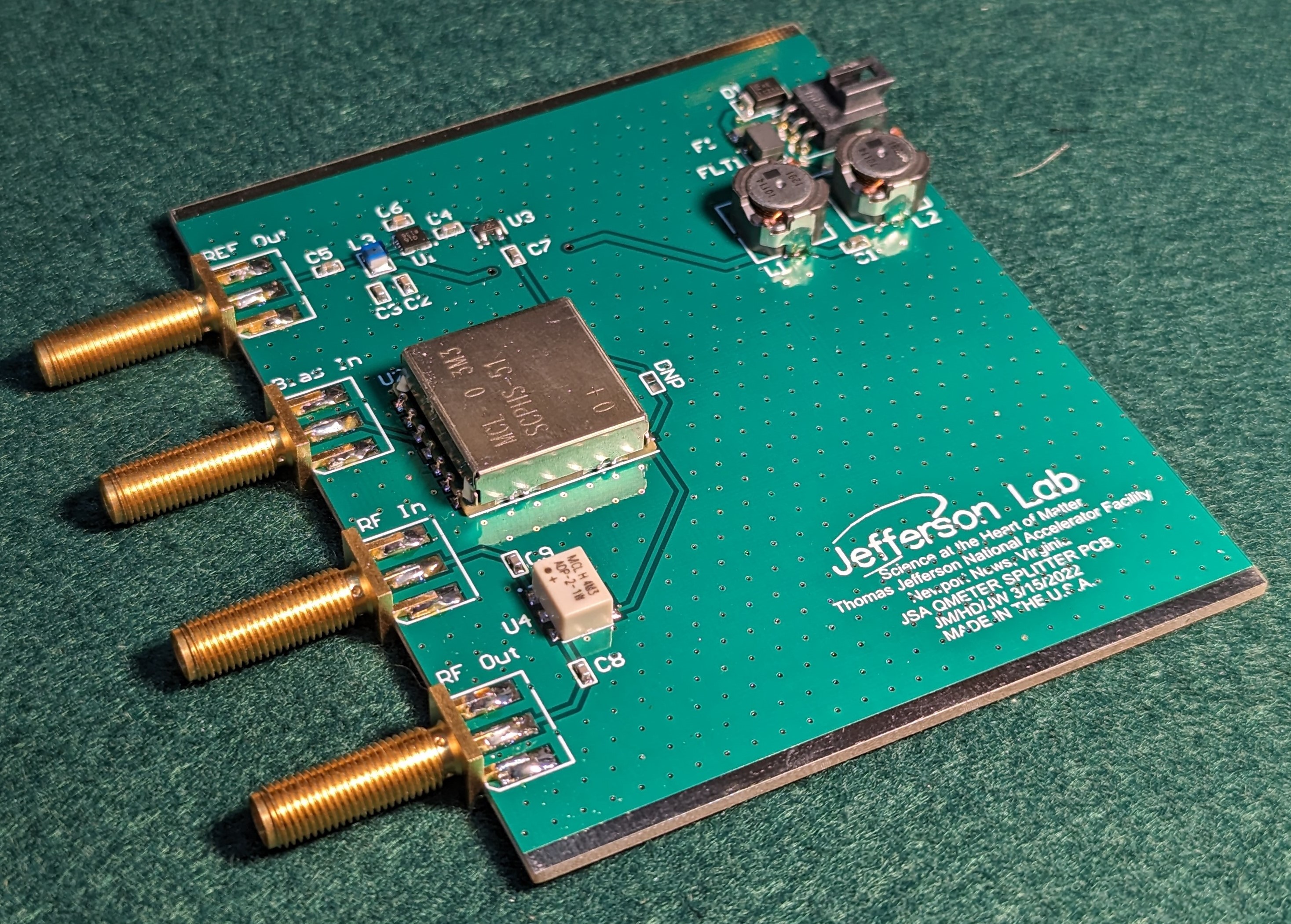}
	\caption{Photograph of the reference conditioner board, with the 360$^\circ$ phase shifter for the deuteron.}
	\label{fig:shifter}
\end{figure}

\subsection{Power supplies}
The Liverpool Q-meter manual specifically discouraged use of CAMAC or NIM crate supplies to avoid ground loops. It used 5 power supplies: 24 and 15\;V for the RF amplifiers, $\pm$15\;V for the LF, and 5\;V for reed relay. Our amplifiers required only $\pm$5\;V, so only a single power supply was necessary. We chose a Condor HAA5-1.5/OVP-AG linear power supply to ensure low noise and jitter. A separate 5\;V supply, the HB5-3-OV-A+G, was used to power the data acquisition system discussed later. Both power supplies were housed within the enclosure, below the mounting plate which held the sub-enclosures seen in Figure \ref{fig:chassis}.

\subsection{Construction}
\begin{figure}
	\centering
		\includegraphics[width= \columnwidth]{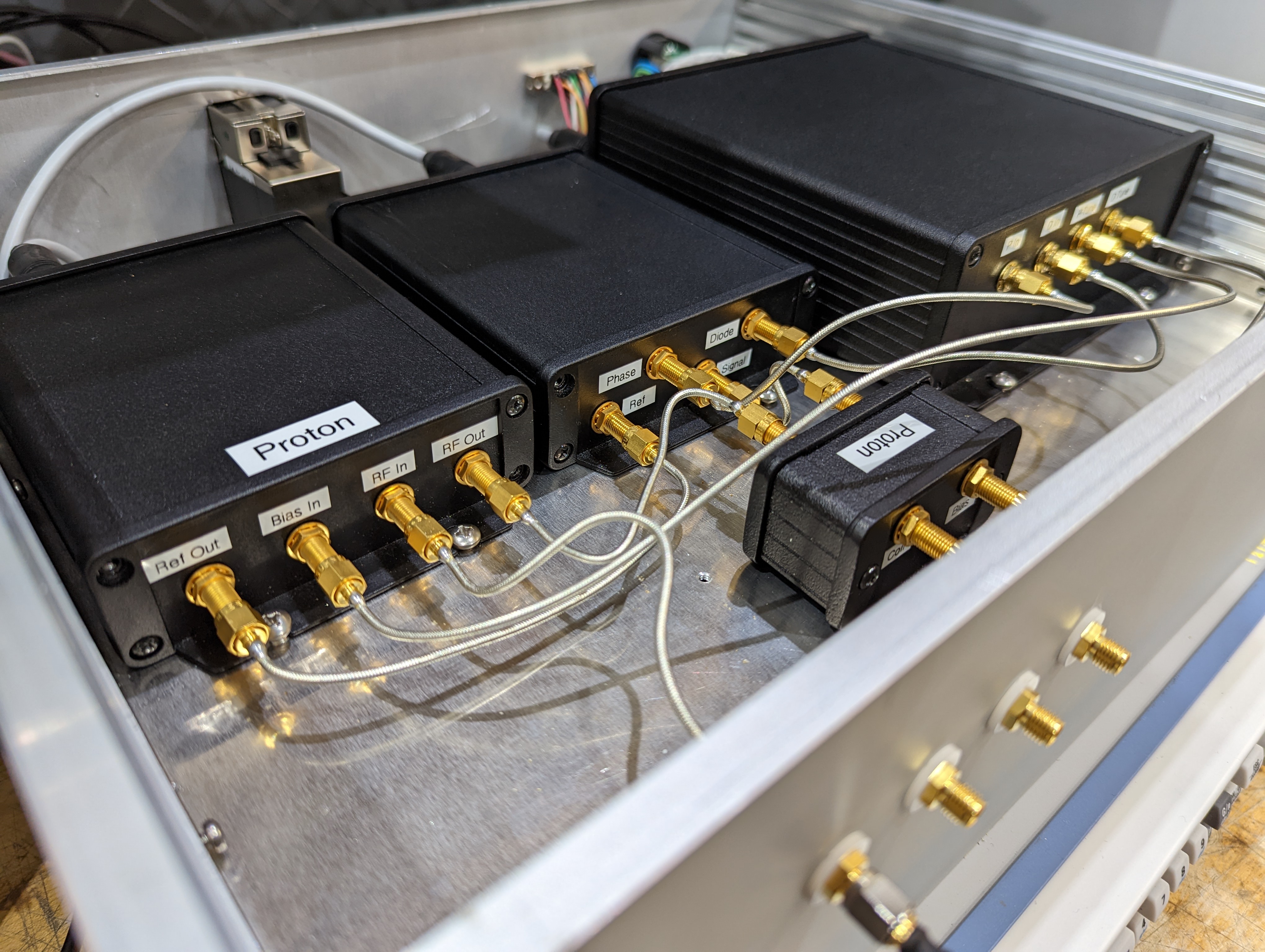}
	\caption{Photograph of the full NMR system chassis with cover removed, showing the four sub-enclosures: reference conditioner, mixer and amplifier, tank and data acquisition.}
	\label{fig:chassis}
\end{figure}

To ease the adoption of our design by other groups, we aimed to use only off-the-shelf components. RF-shield enclosures from Hammond Mfg. were used, as they were provided with conductive end-plate gaskets and easily accommodated Eurocard-style PCBs which slide into extruded slots. The end-plates of these enclosures were machined to accommodate SMA connectors on the front, and M12 power connectors on the back. One enclosure held the mixer and amplifier board assembly, one the reference conditioning board, and a third for the optional warm tank circuit. As seen in Figure \ref{fig:chassis}, these enclosures were mounted on a plate within a 19 inch rack mount chassis, the MultipacPro from nVent Schroff, which included an EMI shielding kit to improve isolation. The power supplies were held within the same chassis, but separated from the Q-meter enclosures below the aluminium mounting plate.

The separation of the Q-meter electronics into three sub-enclosures inside one chassis was designed to allow modularity of parts. The mixer and amplifier boards together in the ``Q-meter'' enclosure form the heart of the system and are limited in frequency by the RF amplifiers from 20 to 500\;MHz.  The signal conditioning board used was specific to the frequency needed, so these boxes could be swapped out to change from a proton to a deuteron measurement. When the tank circuit was used within the chassis, as in Figure \ref{fig:chassis}, only two SMA chassis feed-throughs were used, bringing in the RF from the signal generator and connecting the $\lambda/2$ resonant cable to the coil. When the cryogenic tank was used, all four SMA feedthroughs were needed, adding the supply and varactor bias cables. The SMA connections between components made using other digitization or amplification stages easy to incorporate.

\section{Data Acquisition}
\label{sec:Software}
While the design of our Q-meter followed the Liverpool quite closely, there was significant room for improvement in the control and acquisition electronics that support it. We opted to create a bespoke FPGA-based acquisition system, which we put inside the chassis, seen at top right of Figure \ref{fig:chassis}, preventing the introduction of noise from bringing signals out. This system could communicate via ethernet with any computer on the network, and was controlled using a new Python-based software interface.

The Liverpool Q-meter was often used with Yale cards to provide variable amplification, filtering and voltage offset removal, however we did not require such a system during Run Group C. This is primarily thanks to improvements in ADC resolution; our 24-bit ADCs provide sufficient accuracy at typical output voltages to obviate the need for further amplification. While we have not developed a bespoke device to amplify or remove DC voltage offsets, because our system is connectorized and modular, an SR560 low-noise preamplifier from Stanford Research Systems can be utilized.  This pre-amp has capabilities beyond the Yale card and is available off-the-shelf\,\cite{noauthor_low_nodate}. 

\subsection{Electronics}
\begin{figure}
	\centering
		\includegraphics[width= \columnwidth]{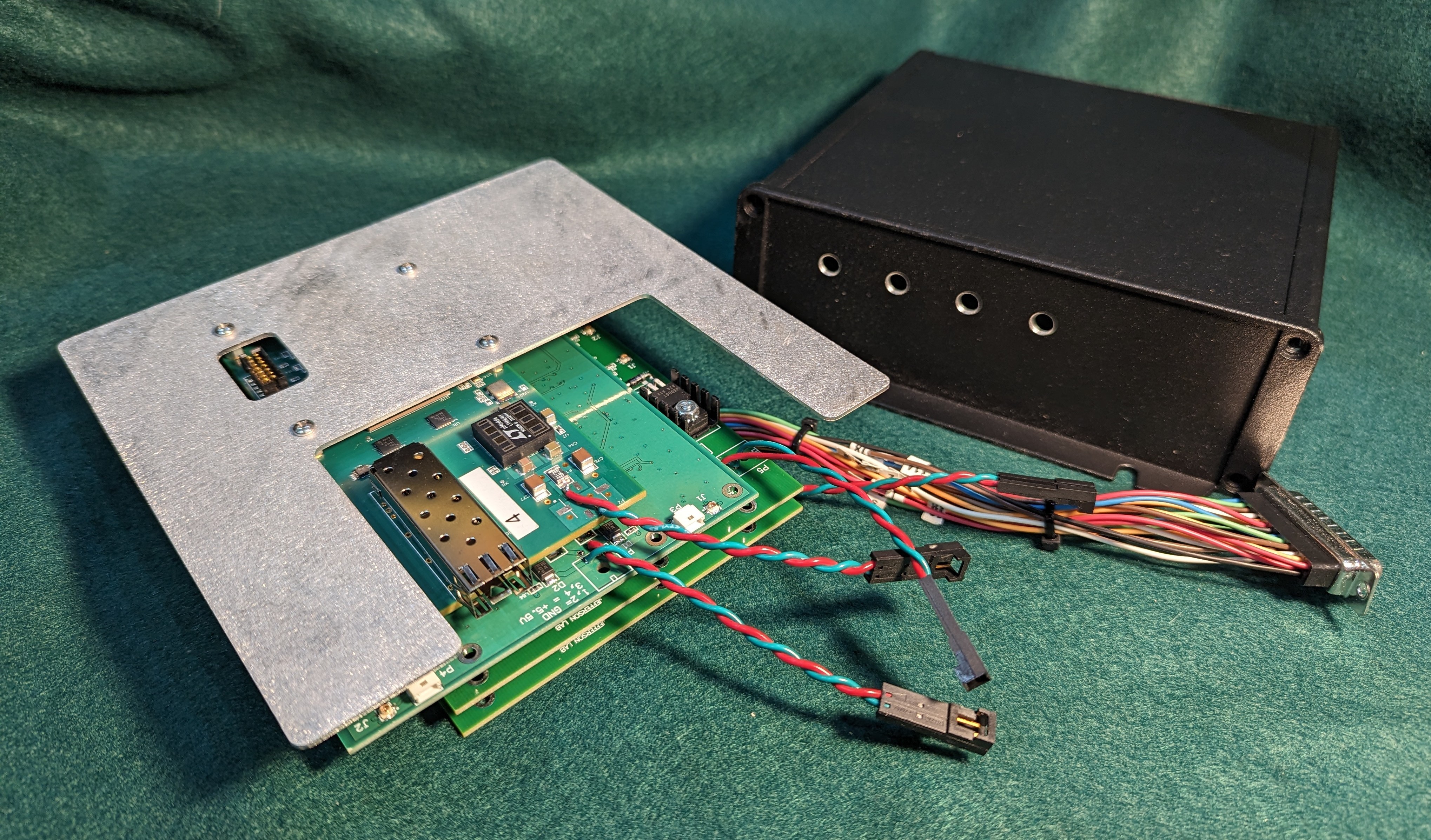}
	\caption{Photograph of the data acquisition hardware. The four boards were hung from an aluminum mounting plate which slides into the enclosure at right.}
	\label{fig:daq}
\end{figure}

This new DAQ system, seen in Figure \ref{fig:daq} had four printed circuit boards which were stacked together within a separate enclosure in the chassis: the control board, a digitization board, a voltage out board and a signal generator control board. The control board used a Xilinx Artix-7 Field Programmable Gate Array (FPGA), which could communicate with an outside host computer via ethernet. The digitization board included two Texas Instruments 24-bit ADS1675 analog-to-digital converters (ADC), to capture the phase and diode signals from the Q-meter. The voltage out board used two Analog Devices 16-bit AD5660 digital-to-analog converters (DAC) to control the varactor variable capacitance on the tank and the electronic phase shifter.

\subsection{Signal Generation}
The final board in the data acquisition enclosure controlled the signal generator that produced the RF supply and its frequency modulation. Previous iterations of Q-meter control at JLab have used a voltage signal from a DAC controlled by a host computer to set the modulation frequency, using a Rohde \& Schwarz generator's external FM Source input. To sweep through frequency, the generator was sent a voltage from -1 to 1\;V to select the lowest to the highest modulution frequency, stepping up and down through a triangle wave, and pausing at each frequency step to allow a measurement. 

However, the Rohde \& Schwarz SMA100A signal generator features an auxiliary input port, allowing faster settling times when changing the frequency. While according to the manual the device is limited to ``less than'' 450\;$\mu$s settle time when switching via list mode, in practice we have found that the external modulation signal can be used for small hops at as fast as 50\;$\mu$s without distortion in the signal. However, the SMA100A's AUX I/O port allows direct connection to the direct digital synthesizer via a 16-bit FM data bus. This permits a very fast frequency hop with settling times of 10\;$\mu$s\,\cite{noauthor_rs_2016}. To take advantage of the 16-bit word data bus on the SMA100A, a fourth board was built to condition 16 digital outputs from the FPGA to drive the signal generator. 

\subsection{Firmware}
We aimed to keep the majority of control and preliminary signal processing on the local FGPA controller, limiting the remote host computer to sending commands to start data collection and receiving collected data.
Custom VHDL firmware was written to operate the Xilinx FPGA Q-meter controller following an algorithm to performed the sweeps through frequency, capturing and sending back the signals. The VHDL had 4 main tasks:  output DAC tuning voltages to control the phase delay and tank capacitance, output the 16-bit word to change the frequency of the Rohde \& Schwarz signal generator,  collect and process data samples from the 2 ADC channels, and send data and receive commands from a host computer via a network interface. The firmware uses the UDP ethernet protocol for commands communication and the TCP protocol to send data to the host computer.

The frequency sweeps are performed many times with their resulting signals averaged to reduce noise. We call each pass back and forth in the full frequency range a \textbf{sweep}, and the full collection of sweeps a \textbf{sweep cycle}, which along with all the other metadata in the system at that time become an \textbf{event}. Typically a cycle contains averaged signals in bunches of between 1000 and 5000 sweeps in events that last between 30 seconds to two minutes. Because these events are long, we send back data to the host computer every few seconds in smaller \textbf{chunks}, which are averaged together at the end of the event.

Before commencing the sweep algorithm in the FPGA, the signal generator must be initialized. This sets the center frequency of the sweep, the modulation width of the frequency sweep (typically 400 kHz), and the RF power output. These settings are done by software at the commencement of each new set of sweeps by direct TCP communication between the host computer and signal generator.

\subsubsection{FPGA Settings}

The FPGA accepts several UDP commands to prepare and control the frequency sweeps. The \textbf{set registers} command sets a number of important parameters to define the sweep behavior and read out, as well as setting the tuning voltages out of the DAC. The \textbf{set} and \textbf{read frequency memory} defines and reads-back the sweep steps that will be undertaken by the algorithm, up to 512 points, allowing the arbitrary distribution of frequency points throughout the modulation range. With this, one could concentrate more data points in a certain region of the sweep, say a 100\;kHz range about center, or completely randomize the order of the frequency points taken.  Finally, commands for \textbf{sweep go} and \textbf{interrupt} control Q-meter operation.

The set registers command requires a number of settings to define the frequency sweep behavior. The \textbf{delay} time sets how long after setting a frequency point to wait, allowing the signal generator to settle. The \textbf{number of samples} determines how many ADC samples to read at each frequency point; this is set as a power of two to allow the averaging of the results bit-wise. The \textbf{total number} of sweeps to take in the sweep cycle must also be set, as well as the \textbf{number of sweeps per chunk} to send back intermittently. The \textbf{ADC configuration} register sets a number of parameters defining the capture mode and rate of the Analog Devices ADC chips.

\subsubsection{Sweep Procedure}

Once the parameter registers have been set, the algorithm begins with the \textbf{sweep go} command. The first frequency memory point is set, and the algorithm pauses the preset amount of time to allow the signal generator to settle on the frequency. Then the two ADCs sample the set number of points from the phase and diode outputs of the Q-meter, performing a bit-wise addition and division to average these together. This average value for each phase and diode channel is summed with the current \textbf{result memory} at the location for the current frequency. Next, the algorithm moves to the next frequency memory point and repeats these steps.

When the algorithm has proceeded up and down through all the frequency points, it assesses how many times it has been through this full sweep. If it has reached the number of sweeps required for a chunk, all the result memory locations are copied to \textbf{read back memory}, and the result memory is reset to zero. With the result memory at zero, the algorithm can continue to sweep to fill the next chunk. In the meantime, the read back memory is sent via TCP back to the host computer. 

The TCPIP packet contains the following data: which chunk this is out of the full cycle, how many sweeps are summed into this chunk, whether this chunk is for phase or diode, and the summed values at each frequency representing the data for this chunk. The host will then divide each summed value by the number of sweeps in the chunk to get an average value at each frequency, which represents the sweep signal for this time period. This continues until the total number of sweeps for this cycle is finished and the host has averaged all the chunks together to finish.

\subsection{Software}
The final piece of this complete refresh of our cw-NMR system was the software to coordinate the operation of the Q-meter, collect and analyze the data, and provide a user interface. Shying away from the proprietary LabVIEW software used previously, the new program was written in Python with a graphical interface based on PyQt5, which handled the full the event-driven interaction flow with threading. In addition to communicating with the FPGA data acquisition over the network, the software also directed communication and control with other crucial components of the DNP system, including microwave frequency tuning and shim magnet power supply. This software added a number of convenience and performance features, drawing on our experience with LabView NMR software used at JLab in the past. The code and documentation are available on GitHub\,\cite{maxwell_jdmaxjlab_pynmr_2021}.

\begin{figure}
	\centering
		\includegraphics[width= \columnwidth]{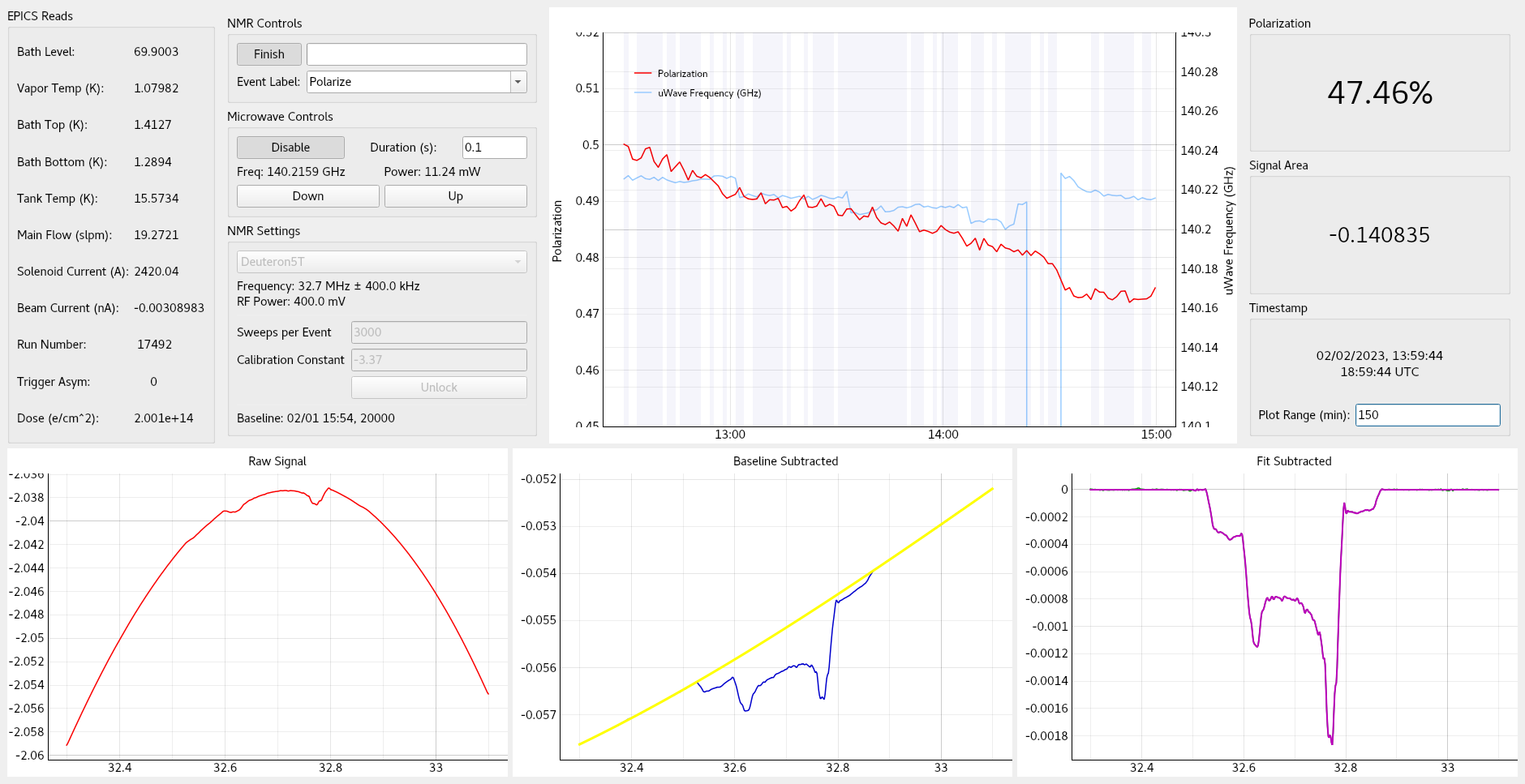}
	\caption{Screen shot of the run tab of the Python control software during the running of Run Group C, with beam on a 47\% polarized ND$_3$ target using the cold tank circuit.}
	\label{fig:run}
\end{figure}

The front end of the interface was split into multiple task-oriented tabs. The \textbf{run tab}, seen in Figure \ref{fig:run}, included the most commonly used controls and displayed plots and indicators showing the status of the NMR system and other systems via JLab's EPICS system\,\cite{white_evolution_1999}. In this tab, settings for the NMR sweeps could be chosen from a drop-down menu, picking a group of settings organized into \textbf{channels} in a YAML configuration file. Pressing the \textbf{run} button would start the sweep algorithm, displaying the chunk data in a raw signal plot as it arrived. At the end of an event, the signal would be analyzed and polarization displayed, meanwhile a new set of sweeps would be started, unless \textbf{pause} was selected. This was an improvement over previous versions, as it allowed more computationally intensive analysis on the host computer, such as curve fitting, while the FPGA continued with the sweep procedure. 

The main plot of the run tab showed the polarization over time, along with the measured microwave frequency applied to the material and the status of the experimental beam---off or on---as a grey background.  This enabled easy identification of the causes of polarization change from either microwave changes or the changes in heat from the beam. Three plots at the bottom showed the three steps of the analysis for the previous event, as chosen in the analysis tab.

A completely new concept, the \textbf{analysis tab} allowed configuration of the analysis of the phase signal in three modular steps as data was being taken. The first step of the analysis requires removal of the Q-curve, the portion of the signal from the Q-meter electronics. This is typically done by taking \textbf{baseline} signals, in which the magnetic field is altered to shift the Larmor frequency of the spins out of the sweep frequency range. In the first modular step of the analysis, a drop-down allowed the choice of a baseline or instead a polynomial fit to the background for subtraction. In the second step of the analysis, any drift in the Q-curve from changes in the system was removed to completely isolate the polarization signal itself.  This was done with a polynomial fit to the selectable ``wings'' of the signal, the portion to the left and right of the curve, excluding the polarization signal itself. 

In the third step the polarization is determined, usually by integrating all the points in the flattened curve, where only the polarization signal should remain. The resulting area is multiplied by the calibration constant to get a polarization for the given event. Other polarization determination methods could be chosen in this step, however, including a simple peak maximum, or a full fit to the deuteron Pake doublet. This Python Pake doublet fitting routine was written in Python, based on C-code by C. Dulya\,\cite{dulya_line-shape_1997}, and is available on GitHub\,\cite{maxwell_jdmaxdeuteronpeakfit_2024}. The results of the analysis choices made in this tab were visible in three plots on the run tab.

\begin{figure}
	\centering
		\includegraphics[width= \columnwidth]{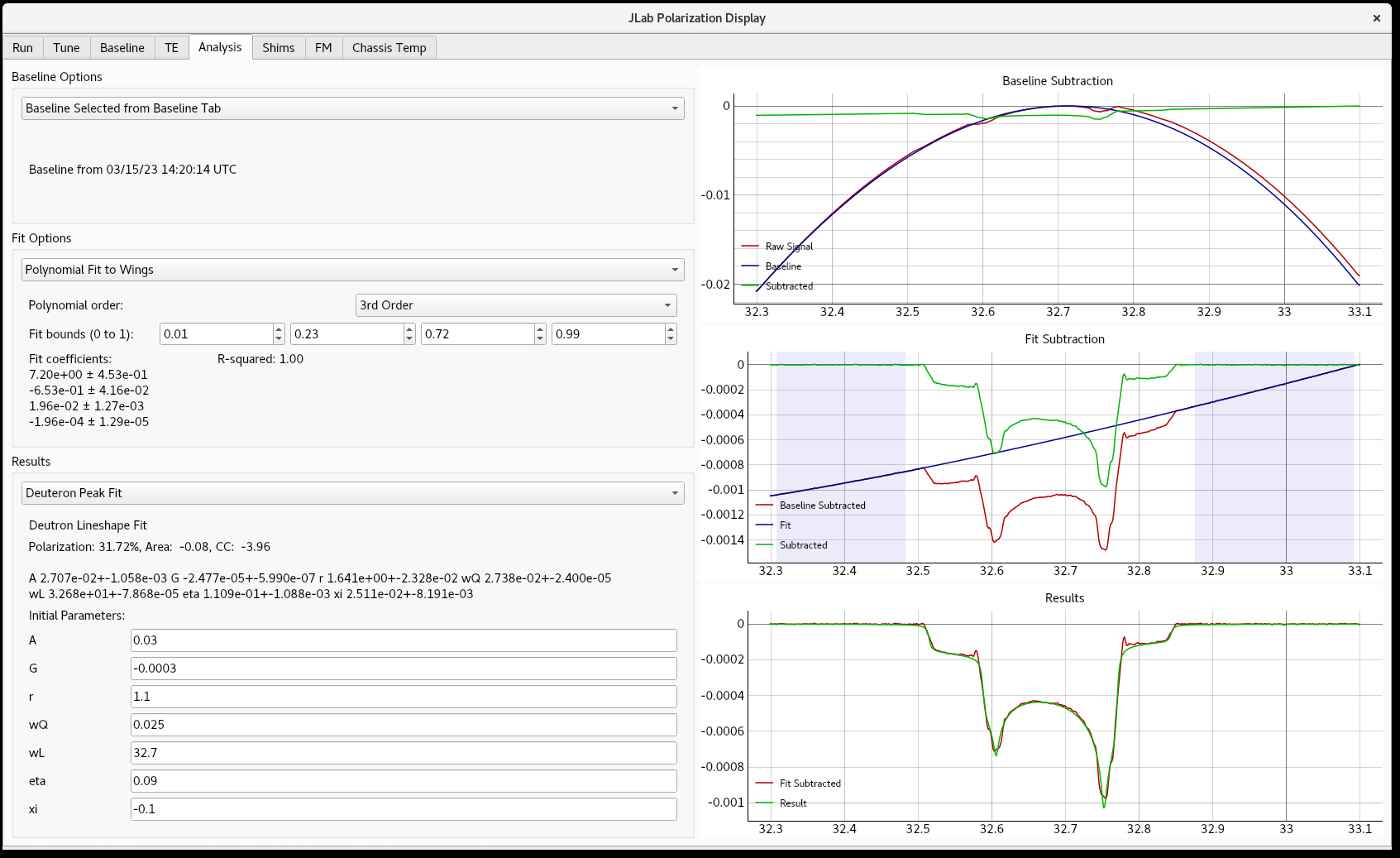}
	\caption{Screen shot of the analysis tab, showing three modular analysis steps: removing the baselines, removing the drift, and fitting the final deuteron line shape to determine the polarization.}
	\label{fig:anal}
\end{figure}

The \textbf{tune tab} was another new addition, taking advantage of the digitization of both the phase and diode signals and the voltage control of phase delay and tank capacitance. The tune tab ran the sweep procedure with very small chunks and very little averaging, displaying the diode and phase Q-curves in real-time along with controls to adjust the tune for each.

The \textbf{baseline tab} allowed the selection of baseline events from previously taken data. To ensure that the chosen baseline curve was well-matched to current running, plots showed the chosen curve and the current running event minus this new baseline. One new addition in this tab is the ability to select multiple events to average together to form a single baseline. The reduction of statistical noise through averaging meant that a baseline should contain at least as many averaged sweeps as the signal is will be subtracted from. By selecting multiple events, more sweeps can be averaged into a single baseline, removing the need to take very long baseline events. 

The \textbf{TE tab} functions were centered around thermal equilibrium measurements. This tab showed plots of the recent event's signal area over time, allowing them to be fit to exponential decay functions and selected for use in a TE measurement. Once the desired points were selected, the calibration constant was calculated and automatically set on the run tab.


\section{Experimental Performance}
\label{sec:Performance}
The first proof of the operation of our final system with polarized material came with the running of Run Group C in Hall B from June 2022 to March 2023. Two Q-meter systems were prepared for this experiment, each with a separate DAQ and signal generator, to allow the measurement of proton and deuteron signals at 5\;T, simultaneously if desired. While the proton Q-meter channel used a tank circuit inside the Q-meter enclosure and a standard $\lambda/2$ cable, the deuteron channel took advantage of a cold tank inside the cryostat (as in Figure \ref{fig:coldboard}) for improved signal-to-noise with the much smaller Pake doublet signal.

Due to experimental constraints in Run Group C, the tank coil was located outside the material cell throughout the experiment, greatly reducing coupling and signal size, and making a more challenging NMR environment than typically encountered at JLab.
While tests of system linearity and circuit performance using resonating crystals showed the viability of the system, we devised a method of direct comparison between the new system and the Liverpool to assess relative performance.

\subsection{Comparison with Liverpool Q-meter}
To allow direct comparison of our new Q-meter system with a Liverpool Q-meter, we utilized a mechanical, single-pole double-throw RF switch, Mini-circuits RC-2SPDT-A18. This switch allowed a single tank circuit and coil to be measured by the Liverpool and JLab Q-meters in turn, without changing other circuit parameters. For this to be possible, it was necessary to a use a Liverpool Q-meter that was modified for an external tank circuit.  In this scheme the supply and signal lines of a external tank circuit were connected to the two channels of the switch, allowing these to be connected to either setup. 

Unfortunately, the addition of the RF switch, splitter and cables necessarily introduced additional noise during these tests, so this setup did not allow a meaningful comparison of the signal-to-noise performance of these systems. Background noise was increased by between 2 and 3 times during these tests. 

The capacitance tuning of the tank for both systems was handled by the new data acquisition, while the phase tuning was separated to the new phase shifter for our Q-meter, and a length of phase delay for the Liverpool. The Liverpool data acquisition was performed using a National Instruments USB-6212 BNC.

\begin{figure}
	\centering
		\includegraphics[width= \columnwidth]{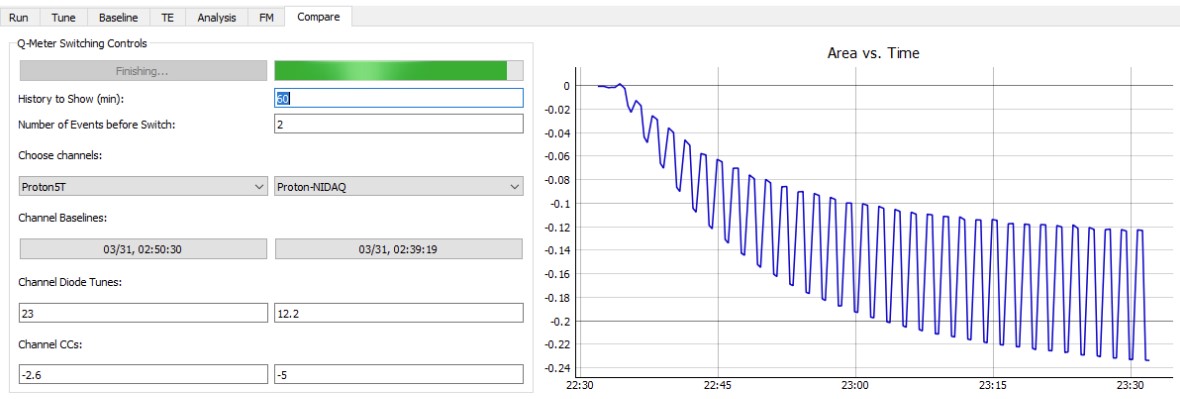}
	\caption{Screen shot of the compare tab alternating measurements with the Liverpool and new Q-meter systems. The channel selection determines which data acquisition system was used, and each channel could use its own baseline, capacitance tune and calibration constant values.}
	\label{fig:compare}
\end{figure}

To make measurements of the same tank circuit alternating between these two systems, the Python software was extended to communicate with the National Instruments DAQ board via NI-DAQmx. A new \textbf{compare tab}, seen in Figure \ref{fig:compare} allowed the selection of any two channels and their respective baselines, diode tune and calibration constant values.

These tests were performed at the end of Run Group C, while the target was still operational but the experimental beam was shut down. While the time available in the hall for these tests was limited by the scheduled removal of the target system, we were able to make the necessary changes to the system to allow the comparison test. We thermalized for calibrations, then polarized protons in TEMPO-doped butanol to roughly 65\% positive and then negative.

\begin{figure}
	\centering
		\includegraphics[width= \columnwidth]{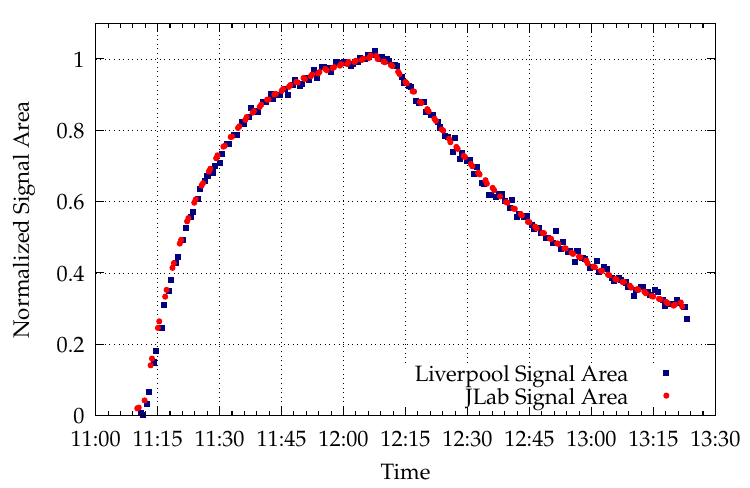}
	\caption{Final signal area for TEMPO-doped butanol during polarization and relaxation. During polarization, the Liverpool and JLab Q-meters alternated data taking on the same tank circuit and material sample. The signal area for both are normalized to the point of highest polarization, which was around 65\%.}
	\label{fig:direct}
\end{figure}

Figure \ref{fig:direct} shows Liverpool and JLab Q-meter polarization signal area during a  polarization and relaxation cycle. Here the area is normalized to the highest point of polarization for each. During the test each system took two measurements, then the RF switch connected the tank to the other system for two measurements. Agreement is very good throughout the dynamic range of the measurement.

\begin{figure}
	\centering
		\includegraphics[width= \columnwidth]{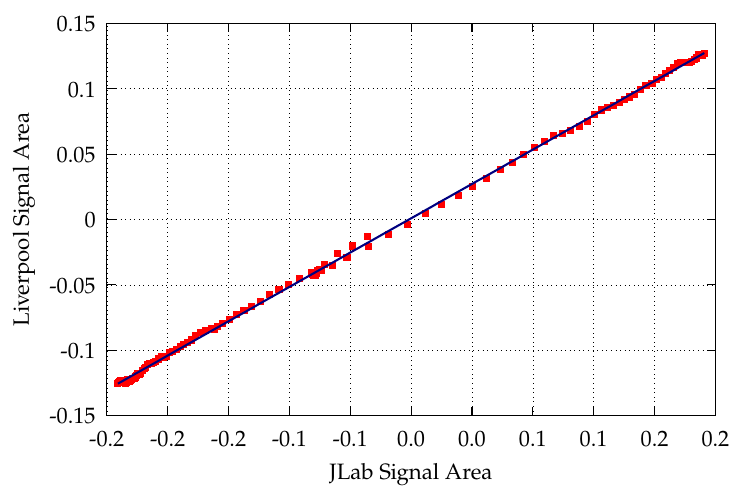}
	\caption{Final signal area from the Liverpool and JLab Q-meters plotted against the other during a cycle of polarizing positively, with negative signal area, and negatively, with positive signal area.}
	\label{fig:linear}
\end{figure}

Figure \ref{fig:linear} shows data from each Q-meter plotted versus the other in a range of polarizations from positive and negative. As the polarization changes quickly, particularly during the pump up or down, pairing data points from each Q-meter based on the nearest in time creates a skew away from linear. To reduce this effect in our comparison,  using a data taking sequence of two Liverpool points followed by two JLab points, we grouped points in four, with two points from each averaged to form a single $(x,y)$ point in this plot. 
So for the series of area measurements $a_i$ in a sequence with JLab measurements at $a_1$, $a_4$, $a_5$, $a_8$, $a_9$ $...$ and Liverpool measurements at  $a_2$, $a_3$, $a_6$, $a_7$, $a_{10}$ $...$ we have:

\begin{equation}
    x_i = \frac{a_i + a_{i+3}}{2}, \quad  y_i = \frac{a_{i+1} + a_{i+2}}{2}
\end{equation}
if $a_i$ is a JLab measurement, and 
\begin{equation}
    x_i = \frac{a_{i+1} + a_{i+2}}{2}, \quad  y_i = \frac{a_{i} + a_{i+3}}{2}
\end{equation}
if $a_i$ is a Liverpool measurement. 

A linear fit to this plot again shows good agreement. Significant deviation from this line at large signal size would indicate a loss in linearity of the device, which is not apparent. Small deviations are seen near zero when the polarization is changing quickly, and at high polarization due to capacitance drifts in time.

\section{Future Efforts}
\label{sec:Future}
With the successful performance of this new Q-meter system in Hall B, we look forward to incorporating a number of improvements which will be facilitated by the modular nature of the device. A quadrature measure of both the dispersive and absorptive protions of the signal would give the complete electronic response of the resonance\,\cite{kisselev_measurement_1995}, and requires measuring both the in-phase and out-of-phase portion of the signal with an additional 90$^{\circ}$ phase shift. This would be best accomplished by replacing the diode chain of the mixer board with a second mixer; the remainder of the amplification and digitization could remain unchanged. 
Electronic tuning allows the synchronous tuning of the capacitance and phase shift to programmatically reduce or even remove the background Q-curve \cite{crabb_proceedings_1998}, and this could be implemented with updates to our VHDL code.  The system's ability to create arbitrary frequency sweep profiles is also yet to be explored. With the flexibility of using tank circuits within the cryostat, we are investigating the addition of amplification on or near the tank circuit to improve signal-to-noise.

While the frequency down-mixing paradigm of the Liverpool Q-meter provides superior performance to the digital lock-in amplifiers we have tested, we are developing a method to allow direct digitization of the tank signal up to 213\;MHz with fast ADCs. While no ADCs exist to densely sample a sine wave above 200\;MHz, our scheme would use variable delays to sample successive cycles at different points, effectively creating an algorithmic down-mix to reconstruct the full wave. A proof of concept  is in development, and the scheme is under patent protection\,\cite{maxwell_digital_2023}. 

Machine learning (ML) techniques are rapidly evolving, and we are collaborating with JLab's experimental Physics software and computing infrastructure group on a method of ML control for polarized targets which includes artificial intelligence for NMR signal processing. One goal is to use AI to improve fits of the Q-curve, which is not quite a low-order polynomial, to better isolate the polarization signal under changing experimental conditions.


\section{Conclusions}
\label{sec:Conclusions}
We have developed and implemented a new Q-meter system in the Liverpool Q-meter style here at Jefferson Lab. This system was produced with off-the-shelf components in a modular layout allowing easy reconfiguration and incorporation of other instruments. The electronic layout was designed to follow the Liverpool Q-meter closely, replacing outdated components for modern counterparts in an all-new board and enclosure scheme. 

While some improvement in signal-to-noise performance can be expected from modern components in the Q-meter itself, it was the supporting electronics of data acquisition and control that offered the most fertile ground for the enhancement of the precision and ergonomics of polarization measurements. Incorporating direct access to the signal generator from the FPGA frequency sweep algorithm allowed an increase in the data taking rate by over three times, and the improvements in digitization resolution meant an added variable amplification device was not needed.

We initially produced two independent Q-meter systems which operated successfully during the running of Hall B's Run Group C from June of 2021 to March of 2022. During this time, 180,000 events representing over half a billion frequency sweeps were performed on polarized NH$_3$ and ND$_3$ targets samples. The new Python software suite allowed live deuteron peak fitting, baseline averaging and TE measurement, facilitating the maximization of polarization and reducing downtime during the experiment. 

A third system is being used a Oak Ridge National lab, and a fourth is in production for use at the University of New Hampshire. In collaboration with ORNL, the proven JLab Q-meter design is being prepared for wider production, and we look forward to sharing these devices for other polarization efforts.

\section*{Acknowledgements}
We gratefully acknowledge the constant support of the Jefferson Lab Target and Fast Electronics Groups. We also thank M.~Houlden and G.~Court, who aided greatly in the simulation of the Q-meter by providing MathCAD calculation notebooks. We remember D.~Crabb, whose lifetime of work in polarized targets inspired and instructed us and so many others.  This material is based on work supported by the U.S.~Department of Energy, Office of Science, Office of Nuclear Physics under contract DE-AC05-06OR23177.

\section*{CRediT authorship contribution statement}
\textbf{J. Maxwell}: Conceptualization, Methodology, Software, Validation, Formal analysis, Investigation, Data Curation, Writing - Original Draft, Visualization, Project administration.
\textbf{J. Brock}: Methodology, Investigation
\textbf{C. Cuevas}:  Conceptualization, Supervision.
\textbf{H. Dong}: Conceptualization, Methodology, Software, Validation, Investigation, Resources.
\textbf{C. Keith}: Conceptualization, Methodology, Supervision, Project administration.
\textbf{J. Pierce}: Conceptualization, Methodology, Supervision.

\bibliographystyle{elsarticle-num}

\end{document}